\documentclass[]{aa}

\def\eps@scaling{.95}
\def\epsscale#1{\gdef\eps@scaling{#1}}

\def\plotone#1{\centering \leavevmode
    \epsfxsize=\eps@scaling\columnwidth \epsfbox{#1}}

\def\plotfiddle#1#2#3#4#5#6#7{\centering \leavevmode
    \vbox to#2{\rule{0pt}{#2}}
    \includegraphics{#1}}

\begin{document}
\thesaurus{03(11.03.4 Virgo; 11.03.1; 11.09.3; 11.19.2; 13.09.2; 13.09.1)}
%\title{Intracluster FIR emission as a tracer of cluster formation}
\title{On the FIR emission from intracluster dust}
\author{Cristina C. Popescu \inst{1,2}
\and  Richard J. Tuffs  \inst1
\and  J\"org Fischera \inst1
\and  Heinrich V\"olk \inst1
}

\offprints{Cristina C. Popescu (email address: popescu@levi.mpi-hd.mpg.de)}
\institute{Max Planck Institut f\"ur Kernphysik, Saupfercheckweg 1, 
           D--69117 Heidelberg, Germany
\and The Astronomical Institute of the Romanian Academy, Str. Cu\c titul de
Argint 5, 75212, Bucharest, Romania}
\date{Received; accepted}
\maketitle
%\markboth{C.C.Popescu et al.:}{}
\begin{abstract}

We make predictions for the diffuse far-infrared (FIR) emission from 
dust in the intracluster medium (ICM) of the Virgo cluster using detailed 
information on potential dust sources, grain heating and sputtering rates 
available for this cluster from recent optical and X-ray studies. 
In the cluster core we identify the winds of red giant and supergiant 
IC stars as the main continuous sources of IC grains, with a 
dust injection rate of 0.17$\,$M$_{\odot}$yr$^{-1}$.
The predicted FIR surface brightness from this 
dust component is however a factor of $\sim\,$10 below the detection limit of currently 
available telescopes.
Grains that are impulsively removed from spiral galaxies by ram-pressure 
stripping as 
they enter the cluster core region can sporadically dominate the
grain injection rate into the ICM. However, these events should lead to the 
appearence of rare,
relatively bright, localised FIR sources around the parent galaxy.

The outer regions of dynamically young clusters like the Virgo cluster 
have a further potential source of intracluster grains since they are still
accreting freshly infalling spiral galaxies which are 
presumably contained in an accreting intergalactic medium (IGM).
We show that cosmics ray driven winds from the infalling spirals can 
inject grains into a subvirial IGM that is external to the observed 
X-ray-emitting ICM.
Sputtering during the injection process and in the IGM is weak, so that
the injected grains should accumulate in the IGM
until the infall brings them into contact with the hot ICM.
Normalising the mass loss rate
in the galactic winds to the mass-loss rate and B-band luminosity of the
Milky Way, we estimate a dust accretion rate of 1.0$\,$M$_{\odot}$yr$^{-1}$
from the infalling IGM . This effect dominates 
the dust injection rate from known sources embedded in the hot Virgo ICM.
Thus, any detection of diffuse IR emission would probe
the current dust accretion rate for the 
cluster, acting as an indicator of the youth and the dynamical state of the 
cluster. The predictions
for the Virgo cluster are generalised to other clusters and 
the possibility of detection of dynamically young clusters at cosmological 
distances is discussed.
Although dominated by the discrete source emission from galactic disks,  
it is possible that diffuse sub-mm dust emission from 
the ICM could be detected in experiments similar to those designed
to map the sub-mm excess 
due to the Sunyaev-Zeldovich effect in distant clusters.
\end{abstract}


\section{Introduction}

Hot gas in clusters can potentially be primordial in origin, heated
adiabatically during the infall into the potential well, or it could 
alternatively be 
hot gas that was ejected from galaxies into the intracluster (IC) medium. 
Studies of X-ray
line emission have revealed surprisingly high metallicities (0.4-0.5),
suggesting that non-negligible or even a substantial fraction of the 
intracluster medium originated in
galaxies. It has also been suggested that the ejected material could 
contain dust and that this dust could produce detectable FIR
emission (Dwek et al. 1990). However, in the hot X-ray emitting plasma from the
centre regions of clusters of galaxies the dust is quite efficiently sputtered 
and destroyed, on time scales of typically a few times 10$^{8}\,$yr. 
To produce a detectable infrared emission this dust has to be injected at a 
high rate at the present epoch. 
Several indirect estimates indicating the presence of a substantial amount of 
grains in the IC medium, sufficient to detect in emission in the infrared, 
have been made. These have been derived from extinction measurements 
(Zwicky 1962; Karachentsev \& Lipovetskii 1969; Bogart \& Wagoner 1973; 
Boyle et al. 1988; Romani \& Maoz 1992), from the soft X-ray absorption 
measurements (Voit \& Donahue 1995, Arnaud \& Mushotzky 1998) or from the 
observed amount of gas, by assuming that all the gas has been 
continuously injected with the 
Galactic gas-to-dust ratio, Z$_{\rm d}=0.0075$ since formation
of the cluster (Dwek et al. 1990).
However, it is not clear whether   
the current populations of elliptical and spiral galaxies seen 
in clusters are capable of supporting the  
required injection rate to achieve detectability of IR emission from the 
IC medium, or whether other sources might be required.

Observational evidence for FIR emission associated with intracluster dust 
has generally been inconclusive (Wise 
et al. 1993). Recently Stickel et al. (1998) detected a colour excess in 
the central 0.2\,Mpc radius of Coma cluster, from the FIR emission measured 
by the ISOPHOT C200 camera aboard ISO. This colour excess was interpreted as 
thermal emission from intracluster dust with a temperature slightly higher 
than in the galactic cirrus and in cluster galaxies. However
Quillen et al. (1999) have argued that the measurement by Stickel et al. 
comes from cluster galaxies. They estimated that the galaxies in the central 
region of the Coma cluster would produce a surface brightness of 
0.06\,MJy/sr, which they take as being in agreement with the Stickel 
detection. Nevertheless, the Stickel et al.
measurement represents only a lower limit and it was obtained after 
subtracting the contribution of foreground galactic cirrus and cluster 
galaxies, on the assumption that the latter emission would arise from cooler 
dust. In summary there are major theoretical and observational uncertainties
concerning both the amount and source(s) of IC dust.

Here we make predictions for intracluster dust emission taking into
account a variety of potential sources of dust. We illustrate the calculations
using data for the Virgo Cluster, which is close enough for any intracluster
IR emission to be spatially distinguished from emission from constituent 
galaxies and has detailed X-ray information, allowing realistic calculations 
of grain heating and sputtering time scales to be made. Furthermore, there is 
detailed information on all the Virgo galaxy members (Binggeli et al. 1993), 
and there has even been a recent detection of an intergalactic
star population (Ferguson et al. 1998), which can be considered as a further
potential source of intracluster grains. Finally, the Virgo cluster is the 
best studied example of a dynamically young cluster, into which
spiral galaxies are falling in from the field (Tully $\&$ Shaya 1984). This
allows a hitherto unconsidered, but potentially dominant, source of IC grains
to be addressed, namely grains embedded in an external subvirial 
intergalactic (IG) medium accreting onto the X-ray emitting 
intracluster medium.

The plan of this paper is as follows: In Sect. 2 we give details of the 
calculation of the infrared emission from stochastically heated
grains in the Virgo IC medium, for grain populations arising from 
a balance between steady state injection and sputtering.
In Sect. 3 we estimate the dust injection rate from potential sources 
(elliptical and spiral galaxies and IC stars) currently seen
in the core of the Virgo cluster. We show that these 
sources are unlikely to provide enough dust to be detected in the IC medium by
current infrared observatories. In Sect. 4 the feasibility of accretion of
grains injected into an external
intergalactic medium from distant infalling spiral galaxies is discussed.
For plausible subvirial gas inflows and physical 
conditions in the intergalactic medium it is concluded that once grains
have been injected into the intergalactic medium, they will descend largely 
unmodified
into the cluster, approximately comoving with the infalling IG medium, and
accumulate over the infall timescale. In Sect. 5 we make detailed
estimates for the efficiency of the grain injection into the IG medium 
by calculating grain sputtering in outflow models for quiescent and star-burst
galaxies. We find that dust grains can survive in winds from quiescent 
(non-starburst) galaxies and that their size distribution is 
not significantly changed by sputtering.
In Sect. 6 we estimate the grain accretion rate from the B-band luminosity
of the infalling population of spiral galaxies, finding this to be the main
source of grains entering the IC medium. Once the
grains reach the hot IC medium they radiate in the 
infrared and are sputtered on timescales smaller than the infall timescale. 
Calculations of the spectrum of the
IR/submillimeter radiation of the accreted grains are given both for grains 
in the outer regions of the diffuse X-ray plateau, and for 
grains directly brought into contact with the dense core region.
Some implications of the results are discussed in Sect. 7.
A distance of 18.2$\,$Mpc to the Virgo cluster is assumed throughout.

\section{Calculation of the infrared emission}

In low-density hot astrophysical plasmas a dust particle is predominantly
heated stochastically by the ambient gas and undergoes temperature 
fluctuations (Gail \& Sedlmayr 1975, Draine \& Anderson 1985, Dwek 1986, 
Dwek \& Arendt 1992). To calculate the temperature distribution for different grain 
sizes in the cluster
we adopted the parameters of the IC gas from 
B\"ohringer et al. (1994), Nulsen \& B\"ohringer (1995) and Schindler et al. 
(1999) based on ROSAT 
observations. The X-ray morphology of the cluster was found to be very similar 
to the structure in the galaxy distribution, with a major component around M87 
and a smaller component around M49. A faint diffuse component was 
also found to trace the cluster out to a distance of $4-5^{\circ}$ from M87 
(also seen by the Ginga satellite; Takano et al. 1989). The diffuse emission
is rather asymmetric, falling off more steeply
to the western side of the cluster. Because of the irregular structure,
previous authors have divided the inner regions of the cluster into
separate spherically symmetrical components centred on M87, M86 and M49,
modelling the X-ray emission of each assuming hydrostatic equilibrium.
Thus 71$\%$ of the total X-ray luminosity  
originates from the M87 halo (out to $1^{\circ}$ from M87) and 
15$\%$ comes from the diffuse X-ray component extending to 
$4-5^{\circ}$ from M87.

\begin{table}[htb]
\caption{}
\begin{tabular}{rccc}
\multicolumn{3}{l}{Density and temperature distribution }\\
\multicolumn{3}{l}{of hot gas in the core of Virgo Cluster}\\
\multicolumn{3}{l}{}\\
\hline\hline
& & \\
radius & n$_{\rm e}$ & T$_{\rm e}$ \\
($^{\prime}$) & (cm$^{-3}$) & (K) \\
& & \\
\hline\hline
& & \\
 0.00-1.67  & 0.0358 & 1.30$\times 10^7$\\
 1.67-3.33  & 0.0149 & 1.61$\times 10^7$\\
 3.33-5.00  & 0.0092 & 2.70$\times 10^7$\\
 5.00-6.67  & 0.0065 & 2.00$\times 10^7$\\
 6.67-7.50  & 0.0058 & 1.69$\times 10^7$\\
 7.50-9.17  & 0.0049 & 2.36$\times 10^7$\\
 9.17-11.25 & 0.0039 & 2.67$\times 10^7$\\
11.25-13.75 & 0.0029 & 3.13$\times 10^7$\\
13.75-16.25 & 0.0023 & 2.38$\times 10^7$\\
16.25-21.25 & 0.0019 & 3.13$\times 10^7$\\
21.25-26.67 & 0.0012 & 1.35$\times 10^8$\\
26.67-37.08 & 0.0008 & 3.71$\times 10^7$\\
37.08-51.25 & 0.0007 & 3.37$\times 10^7$\\  
& &  \\
\hline
\end{tabular}
\end{table}

To calculate grain heating and sputtering rates within the cluster core 
we adopt the deprojected radial density and temperature profiles of Nulsen \&
B\"ohringer (1995), who divided the inner part of the cluster into 38 
concentric spherical shells. Their results are reproduced in Table 1 (where
we have binned together shells for which
the temperature is constant). The abundances were fixed at 0.45 solar
(Koyama et al. 1991; referred to Cosmic values in Allen 1973).
Some recent results based on ASCA observations
(Matsumoto et al. 1996) indicate a higher temperature for the hot plasma in 
the inner 10$^{\prime}$ of the cluster. However, as no deprojected model is 
yet available for 
the new ASCA data, we will use the ROSAT results throughout this
paper. 

To calculate grain heating and sputtering rates appropriate for grains
injected at the perimeter of the diffuse intracluster medium
we estimated the plasma density at a distance of $4-5^{\circ}$ from M87
using the results of B\"ohringer et al. (1994). These authors quote
the total spatially integrated flux of the diffuse component as seen by
ROSAT to be 15$\%$ of the total X-ray flux. 
Assuming spherical symmetry and the broad band emissivity being 
bremsstrahlung dominated, the corresponding average gas density 
is given by:
\begin{eqnarray}
{\rm n}_1^2 = \frac{{\rm F}_1}{{\rm F}_2}\,\frac{\sum{\rm n}_{\rm i}^2\,
{\rm V}_{\rm i}\,{\rm T}_{\rm i}^{1/2}}{{\rm V}_1\,{\rm T}_1^{1/2}} 
\end{eqnarray}
\\
where n$_{1}$, V$_{1}$, F$_{1}$ and T$_{1}$ are the density, volume, flux and 
temperature of the diffuse X-ray component of the cluster and n$_{\rm i}$, 
V$_{\rm i}$, F$_{\rm i}$ and T$_{\rm i}$ are
the same quantities, associated with the spherical shells within the halo of 
M87 (see Table 1). Since 
the temperature remains constant outside the halo of M87 with a 
temperature of $\sim\,$3$\,$x$\,$10$^{7}\,$K (B\"ohringer et al. 
1994), we obtain an average gas density for the diffuse component of 
$4\times 10^{-5}\,{\rm cm}^{-3}$. This is comparable with the gas density
at the edge of the diffuse emission region
which can be derived from Schindler et al. (1999), for a model where the 
gas is in hydrostatic equilibrium.  

In the hot intracluster medium the grains are heated due to inelastic
collisions with electrons and ions. In characterising the dust properties we 
considered spherical 
\lq\lq astronomical silicate\rq\rq\, grains and heat
capacities from Guha-thakurta \& Draine (1989), derived as a 
fit to experimental results for SiO$_2$ and obsidian at temperatures $10< 
{\rm T} < 300$\,K (Leger et al. 1985), with a simple extrapolation 
for ${\rm T}>300$\,K. In Sect. 6 we also considered graphite grains, with 
heat capacities taken from Dwek (1986). The absorption efficiencies 
Q$_{\nu}$(a) were taken from
Laor \& Draine (1993) with grain sizes in the size interval 
[${\rm a}_{\rm min},{\rm a}_{\rm max}$] from 10\,${\rm \AA}$ to 
0.25\,${\mu}$m.  

For the heating of the grains in a hot plasma we adopted the method used
by Dwek (1987). Here the grain is assumed to have an effective thickness 
${\rm R}_0=4{\rm a}/3$ and let R(E) be the range at which gas particles with 
kinetic energy E would be stopped. If R(E) is shorter than  the effective 
thickness, then all energy will go onto the grain. Otherwise the gas 
particles will have a rest energy E', that is given by ${\rm R(E')=R(E)-R}_0$. 

In the case of electron heating we used an analytical expression of
the electron stopping power R(E)$\rho$ for silicate and graphite 
from Dwek \& Smith (1996). In the case of ion heating the energy fraction 
deposited by ions onto the grains was calculated with a formula
similar to that used by Dwek \& Werner (1981), which is based on an 
approximation from Draine \& Salpeter (1979), for the range of H and He in 
solids up to 100 keV:
\begin{eqnarray}
  \zeta({\rm E})=\left\{
	\begin{array}{ll}
		1 &, \quad {\rm E}< {\rm E^*},\\
		{\rm E^*/E} &, \quad {\rm otherwise},\\
	\end{array}
	\right.
\end{eqnarray}
\\
where E$^*$ is the minimum kinetic energy of a nucleus to penetrate the
grain. It is given by
\begin{eqnarray}
  {\rm E^*[keV]} = \frac{1}{3}~{\rm a}[\mu m]~\rho[{\rm g/cm^3}]\times \\ \nonumber
	\left\{
	\begin{array}{ll}
		133 & \quad {\rm for~H~atoms}, \\
		222 & \quad {\rm for~He~atoms}, \\
		665 & \quad {\rm for~C,~N,~O~atoms}. \\
	\end{array}
	\right.
\end{eqnarray}
\\
This shows that in a hot gas the ion heating of very small grains is nearly
discrete. 

\begin{figure}[htb]
\plotfiddle{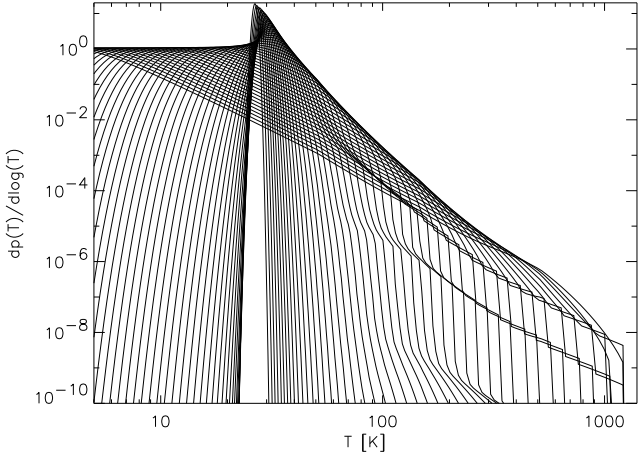}{3.0in}{90.}{37.}{37.}{150}{-20}
\caption[]{The temperature distribution for various silicate grain sizes in the
centre region of 1$^{\prime}.67$ radius of the Virgo Cluster. The grain sizes
are chosen in the size interval from 10\,${\rm \AA}$ to 0.25\,${\mu}$m, with a
logarithmic step size of 0.05. The electron
density and temperature in this region are given in Table 1.}
\end{figure} 

In low density plasma, like that existing in the intracluster medium, the
grains are mainly stochastically heated. The stochastic heating processes are 
calculated following the method of Guhathakurta \& Draine (1989). This method
derives the temperature distribution P(a,T$_{\rm d}$)
of various grain radii {\bf a} as a function of dust temperature T$_{\rm d}$.

In Fig. 1 we give the temperature distribution P(a,T$_{\rm d}$) for various
silicate grain sizes, for the inner 
1$^{\prime}.67$ radius of 
the Virgo Cluster (n$_{\rm e}=0.0358$ ${\rm cm}^{-3}$, 
T$_{\rm e}=1.30\times 10^7$).
Due to stochastic heating, small grains undergo significant 
fluctuations from the equilibrium temperature, while larger grains have
probability functions becoming narrower, eventually approaching delta 
functions. Many of the smaller grains exhibit a plateau in their probability
distribution, separated by steps corresponding to the energy thresholds at
which the various ions are stopped by the grains (see Eqs. (2) and (3)). 

In order to calculate the contribution of all grain sizes we have to derive the
grain size distribution. We consider that the grains are continuously
injected in the IC medium with a grain size distribution given by a power law,
a$^{-k}$.  
The mechanism through which they can be injected is
discussed in the next sections; here we consider the general case. The
grain size distribution can be determined approximately by balancing the local
rate of grain injection into the ICM, S(a), and the rate of grain 
destruction by sputtering:

\begin{eqnarray}
\frac{\partial}{\partial {\rm a}} \left ({\rm N}({\rm a})\, \frac{{\rm da}}{{\rm dt}} 
\right) = {\rm S}({\rm a})
%\frac{{\rm dN}({\rm a})}{{\rm da}} \sim \frac{{\rm t}_{sput}({\rm a})}{{\rm a}}\,
%{\rm a}^{-k}
\end{eqnarray}
\\
where N(a)\,da is the total number of particles with sizes in the interval 
[a,a+da].

For gas temperatures $10^6 < {\rm T} < 10^9$\,K the sputtering timescale for a
silicate or graphite dust particle of radius $a$ is given by Draine \& 
Salpeter (1979):

\begin{eqnarray}
 {\rm t}_{\rm sput}[{\rm yr}] \approx \frac{10^6\,{\rm a}[{\mu}{\rm m}]}
{{\rm n}_{\rm H}[{\rm cm}^{-3}]}
\end{eqnarray}
\\
where n$_{\rm H}$ is the gas density.

From Eqs. (4) and (5) results the steady state grain size distribution:

\begin{eqnarray}
{\rm N}({\rm a}) = \frac{{\rm t}_{\rm sput}}{\rm a}\,
\frac{{\rm dM}_{\rm i}}{{\rm dt}}\,
\frac{{\rm a}^{-k+1} - {\rm a}_{\rm max}^{-k+1}}{{\rm k-1}} \times \\ \nonumber
\times \left [ \int_{a_{\rm min}}^{a_{\rm max}} {\rm a}^{-k}\,
\frac{4\,{\pi}\,{\rm a}^3}{3}\,{\rho}\,{\rm da} \right ] ^{-1}
\end{eqnarray}
\\
where ${\rho}$ is the density of the 
dust grains and ${\rm dM}_{\rm i}/\!{\rm dt}$ is the total dust injection
rate:

\begin{eqnarray}
\frac{\rm dM_i}{\rm dt} = \int_{a_{\rm min}}^{a_{\rm max}} {\rm S(a)}\,
\frac{4\,{\pi}\,{\rm a}^3}{3}\,{\rho}\,{\rm da}
\end{eqnarray}
\\

The total dust mass is then given by:

%\begin{eqnarray}
%{\rm M}=\frac{\displaystyle\frac{10^6}{{\rm n}_H}\,\frac{{\rm dM}_i}{dt}
%\frac{1}{{\rm k-1}}}{\displaystyle \int_{a_{min}}^{a_{max}}{\rm a}^{-k}\,{\rm 
%a}^3\,{\rm da}}\int_{a_{min}}^{a_{max}} {\rm a}^{-k+1}\,{\rm a}^3\,{\rm da}
%\end{eqnarray}\\\\
\begin{eqnarray}
{\rm M}=\int_{a_{\rm min}}^{a_{\rm max}}{\rm N}({\rm a})\,\frac{4\,{\pi}\,
{\rm a}^3}{3}\,{\rho}\,{\rm da}
\end{eqnarray}
\\
and the total infrared emission is:

\begin{eqnarray}
%{\rm F}_{\nu} & = & \frac{\displaystyle \frac{10^6} {{\rm n}_{\rm H}}\,\frac
%{{\rm dM}_{\rm i}}{{\rm dt}} \frac{1} {{\rm k-1}} \frac{1} {{\rm d}^2}}
%{\displaystyle \int_{a_{\rm min}}^{a_{\rm max}}{\rm a}^{-k}\,\rho\,\frac{4\,
%{\pi}\,%{\rm a}^3}{3}\,{\rm da}}\int_{a_{\rm min}}^{a_{\rm max}} 
%{\rm a}^{-k+1}\,{\rm da}\,%{\pi}\,{\rm a}^2\,{\rm Q}_{\nu}({\rm a})\times  
%\nonumber \\%& & \times\int_0^{\infty}{\rm B}_{\nu}({\rm T}_{\rm d})\,
%{\rm P}({\rm a},%{\rm T}_{\rm d})\,{\rm dT}_{\rm d}
{\rm F}_{\nu} & = & \frac{1} {{\rm d}^2}\int_{a_{\rm min}}^{a_{\rm max}} 
{\rm N}({\rm a})\,{\rm da}\,{\pi}\,{\rm a}^2\,{\rm Q}_{\nu}({\rm a})
\times  \nonumber \\
& & \times\int_0^{\infty}{\rm B}_{\nu}({\rm T}_{\rm d})\,{\rm P}({\rm a},
{\rm T}_{\rm d})\,{\rm dT}_{\rm d}  
\end{eqnarray}
\\
where B$_{\nu}$ is the Planck function.

\section{Predicted FIR emission from intracluster dust produced by sources 
inside the core region of the Virgo cluster (within $1^{\circ}$ of M87)}

\subsection{Dust originating in individual intergalactic stars}

In this section we estimate the amount of dust produced through stellar winds
by individual intergalactic red-giant-branch stars, as
detected recently in the Virgo cluster by Ferguson et al. (1998). The
motivation to consider these stars as potential sources of dust
was the suggestion that the intergalactic stars  are 
likely to have originated primarily from the elliptical and S0 galaxies, and
thus to contain a higher proportion of M supergiants to giants as compared to 
the Galactic disk, and therefore a higher integrated mass-loss and dust 
production. Furthermore, this dust would not suffer sputtering 
losses in the injection process, as is the case for
injection from galaxies into the intracluster medium through galactic winds.
According to Ferguson et al. the 
relatively smooth distribution of mass inferred from the X-ray observations 
suggests that most of the intergalactic material was stripped by tidal 
interactions with the cluster potential. Since early-type galaxies are more 
numerous in the cluster, have older stellar populations, and are likely to 
have inhabited the central megaparsec of the cluster for much longer than the 
spiral and irregular galaxies, we will consider here that the
intergalactic stars were stripped from early type galaxies. Nevertheless,
the possibility that some 
of the stars formed in situ, or that some were stripped off by impulsive 
interactions between galaxies (\lq harassment\rq, Moore et al. 1996) cannot 
be ruled out (Ferguson et al. 1998).

The observations made with the Hubble Space Telescope WFPC2 camera with the
F814W (approximately I-band) filter on a blank field in the Virgo cluster 
(Ferguson et al. 1998) indicated a clear excess 
in source counts in the Virgo cluster image. The total flux from the excess 
sources was used by Ferguson et al. to calculate 
the total mass of stars below their detection limits. For this they
considered a population with a metallicity, expressed as a decimal logarithm
relative to the
solar iron-to-hydrogen ratio, [Fe/H]=-0.7, an age of 13 Gyr, a Salpeter 
initial mass function, and a distance of 18.2\,Mpc. They derived an 
underlying surface mass density of
0.14 M$_{\odot}$\,${\rm pc}^{-2}$ if the initial mass function 
continues to 
0.1\,${\rm M}_{\odot}$. 

To calculate the total mass loss of all the stars within the X-ray
emitting core region of 60 arcmin radius from the position of M87, we consider that
the stars follow the distribution of elliptical (E) and S0 galaxies. Their
distribution is described by a spherical King model:

\begin{eqnarray}
{\rho}(\xi) = \rho_0\,(1+\xi^2)^{-3/2}
\end{eqnarray}\\
where ${\rho}_0$ is the central mass density of stars, 
${\xi}={\rm r}/{\rm r}_{\rm c}$ is the
radius in units of the core radius ${\rm r}_{\rm c}$. We take the core radius
${\rm r}_{\rm c}=1.1^{\circ}$, as given for the E+S0 Virgo galaxies by 
Binggeli et al. (1987). The projected mass surface density of stars is 
then:

\begin{eqnarray}
{\Sigma} ({\lambda}) = \frac{2\,{\rho}_0\,{\rm r}_{\rm c}} {(1+{\lambda}^2)}\,
\left( 1-\frac{1+{\lambda}^2} {1+{\xi}_{\rm t}^2}\right) ^{1/2}  
\end{eqnarray}\\
where ${\lambda}={\rm R}/{\rm r}_{\rm c}$ is the projected radius in units of the 
core radius and
${\xi}_{\rm t} = {\rm r}_{\rm t}/{\rm r}_{\rm c}$ is the radius of the outer 
boundary in units of core
radius. If we consider a surface mass density of intracluster stars of 
0.14$\,{\rm M}_{\odot}\,{\rm pc}^{-2}$, as derived by Ferguson et al. (1998) at
a distance of 44.5 arcmin from M87, from Eq. (10) we obtain the central mass 
density of stars, ${\rho}_0 = 0.363\times 10^{-6}\,{\rm M}_{\odot}\,
{\rm pc}^{-3}$.

\begin{table}[htb]
\caption[] {The distribution of stellar masses and dust injection 
 rates in the central region  of the Virgo cluster.}
\begin{tabular}{rcc}
\multicolumn{3}{l}{{\bf}}\\
\hline\hline
& &  \\
1      & 2          & 3  \\
r & M$_{\rm stars}$ &  ${\dot{{\rm M}}}_{\rm i}$ \\
($^{\prime}$) & (M$_{\odot}$) & (M$_{\odot}\, {\rm yr}^{-1}$)\\
& & \\
\hline\hline
& & \\
  1.67 &  1.05e+06 & 4.96e-06\\	
  3.33 &  7.26e+06 & 3.93e-05\\	
  5.00 &  1.98e+07 & 1.33e-04\\	
  6.67 &  3.83e+07 & 3.14e-04\\	
  7.50 &  2.78e+07 & 4.45e-04\\	
  9.17 &  7.69e+07 & 8.08e-04\\	
 11.25 &  1.42e+08 & 1.48e-03\\	
 13.75 &  2.51e+08 & 2.66e-03\\	
 16.25 &  3.54e+08 & 4.34e-03\\	
 21.25 &  1.06e+09 & 9.34e-03\\	
 26.67 &  1.75e+09 & 1.76e-02\\	
 37.08 &  5.23e+09 & 4.23e-02\\	
 51.25 &  1.07e+10 & 9.29e-02\\
\hline
&&\\
Total & 2.0e+10 & 0.17 \\	 
& & \\
\hline
\end{tabular}
\end{table}

The stellar mass within radius ${\xi}$ in units of the stellar core mass
M$_{\rm c}$
 is:

\begin{eqnarray}
{\rm M}(\xi)=3[\ln(\xi+(1+{\xi}^2)^{1/2})-\xi(1+{\xi}^2)^{-1/2}]
\end{eqnarray}\\
with the core mass

\begin{eqnarray}
{\rm M}_{\rm c}=\frac{4\,\pi\,{\rm r}_{\rm c}^3\,{\rho}_0}{3}
\end{eqnarray}\\
From Eqs. (10) and (11) we calculate the stellar mass in each of the spherical shells
from Table 1 and the results are given in Table 2, Column 2. In order to 
obtain 
the number of stars of different 
spectral types and luminosity classes that correspond to the derived total 
mass we have to integrate over the mass function of such a population. We 
consider a model HRD population for an elliptical galaxy assuming the same 
parameters as used by Ferguson et al. (1998) to derive the total mass of the 
stars. For the mass-loss rates and gas to dust ratio in the outflow of 
different types of stars we took the values from Whittet (1992) and Gehrz 
(1989) for the Galactic stellar population. Whilst bearing in mind that the 
averaged mass-loss and dust to gas ratio are subject to considerable 
uncertainty, we may draw some general conclusions from these data. Summing 
individual injection rates for stardust we obtain a total injection 
rate of dust ${\dot{{\rm M}}}_{\rm i}$ for each
spherical shell, and the corresponding values are given in Table 2, Column 3. 
The main contribution to the injection rate comes from M supergiants, which 
would imply that most of the grains should consist of silicates. 

\begin{figure}[htb]
\plotfiddle{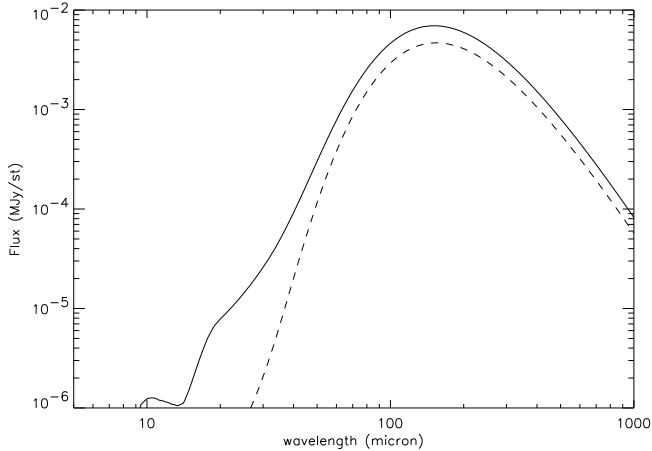}{3.0in}{0.}{50.}{50.}{-150}{-160}
\caption[]{The integrated line-of-sight spectrum in the centre of Virgo 
cluster, given for different sources of dust inside the X-ray emitting core
region: solid line for stars and dashed-line for elliptical galaxies. A
silicate dust composition was assumed. }
\end{figure}

\begin{figure}[htb]
\plotfiddle{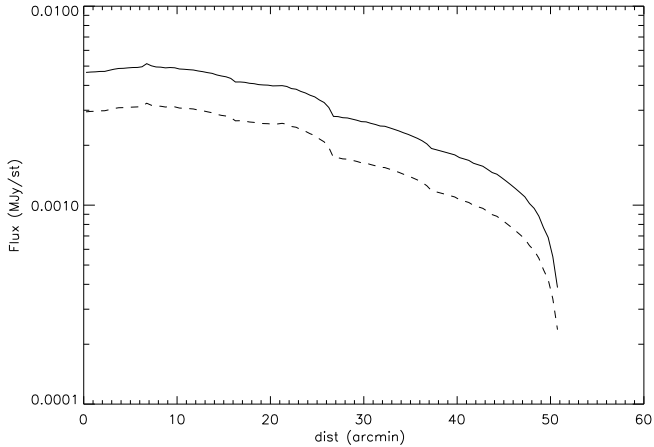}{3.0in}{0.}{50.}{50.}{-150}{-160}
\caption[]{The infrared brightness profile at 100\,$\mu$m for different 
sources of dust inside the X-ray emitting core region: solid line for stars
 and dashed-line for elliptical galaxies. A
silicate dust composition was assumed. }
\end{figure}

The values of the dust injection rate are quite low, indicating 
that there are
not enough stars to produce a detectable amount of dust in the inner hot region
of the cluster, since the grains are sputtered very efficiently by the ambient
hot gas. 

We assume that the grain size distribution injected into the IC medium by the
stars is given by a power law with an exponent given by the MRN (Mathis et al. 
1977) value of k=3.5. We also assume a silicate dust 
composition. From Eqs. (8) and (9) we calculated the infrared
intensity at different projected distances from 
the cluster by integrating over
the line of sight through the cluster. The corresponding infrared spectrum 
in the inner 1$^{\prime}.67$ region of the Virgo cluster is plotted with a
solid line in Fig. 2, while in Fig. 3 we give the radial brightness profile at
100\,${\mu}$m. The infrared emission from dust injected by intergalactic stars
inside the core region of the Virgo cluster is less than 0.01 MJy/st, about
an order of magnitude below the detection limit of $\sim\,$0.1\,MJy/st for 
currently available telescopes. 

\subsection{Dust originating in early type galaxies} 

\begin{table}[htb]
\caption[] {Upper limits for the distribution of dust injection rates 
produced by elliptical galaxies in the central region  of the Virgo cluster.}
\begin{tabular}{rc}
\multicolumn{2}{l}{{\bf}}\\
\hline\hline
& \\
1      & 2 \\
r &  ${\dot{{\rm M}}}_{\rm i}$ \\
($^{\prime}$) & (M$_{\odot}\, {\rm yr}^{-1}$)\\
& \\
\hline\hline
& \\
  1.67  & 1.19e-06\\	
  3.33  & 9.46e-06\\	
  5.00  & 3.20e-05\\	
  6.67  & 7.56e-05\\	
  7.50  & 1.07e-04\\	
  9.17  & 1.95e-04\\	
 11.25  & 3.57e-04\\	
 13.75  & 6.43e-04\\	
 16.25  & 1.05e-03\\	
 21.25  & 2.26e-03\\	
 26.67  & 4.25e-03\\	
 37.08  & 1.02e-02\\	
 51.25  & 2.24e-02\\	
\hline
& \\
Total & 4.15e-02 \\	 
& \\
\hline
\end{tabular}
\end{table}

%\begin{table}[htb]
%\caption[] {The distribution of late-type galaxy blue luminosities and the 
%dust injection rates produced by these galaxies in the central region  
%of the Virgo cluster.}
%\begin{tabular}{rcc}
%\multicolumn{3}{l}{{\bf}}\\
%\hline\hline
%& &  \\
%1      & 2          & 3  \\
%r & L$_{B}$ &  ${\dot{{\rm M}}}_i$ \\
%($^{\prime}$) & (L$_{\odot}$) & (M$_{\odot}\, {\rm yr}^{-1}$)\\
%& & \\
%\hline\hline
%& & \\
%  1.67 & 1.72e+05  & 1.29e-07\\	
%  3.33 & 2.28e+06  & 1.71e-06\\	
%  5.00 & 5.98e+06  & 4.49e-06\\	
%  6.67 & 1.12e+07  & 8.39e-06\\	
%  7.50 & 8.43e+06  & 6.32e-06\\	
%  9.17 & 2.31e+07  & 1.73e-05\\	
% 11.25 & 4.34e+07  & 3.25e-05\\	
% 13.75 & 7.76e+07  & 5.82e-05\\	
% 16.25 & 1.12e+08  & 8.36e-05\\	
% 21.25 & 3.48e+08  & 2.61e-04\\	
% 26.67 & 6.09e+08  & 4.57e-04\\	
% 37.08 & 2.04e+09  & 1.53e-03\\	
% 51.25 & 5.13e+09  & 3.85e-03\\	 
%& & \\
%\hline
%\end{tabular}
%\end{table}

The elliptical and S0 galaxies are not only the best candidates for the 
sources of the intracluster stars (discussed in the previous section), but 
they are also thought to release gas  
into the ICM via supernovae-driven galactic winds. Because these
galaxies have remained in the inner region of the cluster since their
formation, their integrated mass-loss has been used 
(Okazaki et al. 1993) to calculate the galaxian contribution to the 
observed amount of gas in the centre of the cluster. Okazaki
et al. selected all E, S0 and dwarf elliptical galaxies from the Virgo Cluster
Catalogue of Binggeli et al. (1985), in the field within
3$^{\circ}$ from M87. They used three models for elliptical galaxy 
formation with galactic winds (Arimoto \& Yoshii 1987, Matteucci \& 
Tornambe 1987, David et al. 1990,1991a,b) and estimated the masses of 
gas that have been ejected from all these galaxies during the 
cluster lifetime. The result was that even the largest mass obtained by one 
of their model prediction is only 10$\%$ of the value derived from X-ray
observations. They
concluded that elliptical galaxies cannot produce the observed amount
of gas and that 90$\%$ of the gas must be of primordial origin. In the case of
the dust released by elliptical galaxies via supernovae-driven galactic
winds, it is obvious that only the dust produced quite recently
can give rise to an IR emission, since the dust is quickly sputtered
away. But elliptical galaxies have released most of their gas in the earlier
epochs of the cluster lifetime, so their current injection rate is low. We 
will show that even for the most optimistic assumption, of a steady state 
mass-loss over the cluster lifetime, only a
negligible amount of dust is predicted, since the elliptical galaxies are 
well depleted of dust
in comparison with the spiral galaxies (Tsai \& Mathews 1996). Thus if we 
take the amount of gas predicted by Okazaki et al., an average dust-to-gas 
ratio a hundred times lower than the Galactic value (Tsai \& Mathews 1996) 
and if we divide this value by the cluster lifetime we obtain an
average dust injection rate which is an upper limit for the present dust 
injection rate. Tsai \& Mathews (1995) have shown that in the elliptical 
galaxies the grain size distribution is given by a power law with index 
$({\rm k}-1)$. They have also shown that this solution
is valid when the sputtering time is short compared to the radial flow time,
but the same solution is still an excellent approximation even when the
sputtering time is so long that the grains can move inwards during their
lifetime across an appreciable part of the galaxy. Assuming again that the
grains in the elliptical galaxies are injected from the evolving stars with a
power law having the MRN index ${\rm k}=3.5$, the size distribution of the 
ellipticals
will be a power law of index 2.5. We also assume that the 
grain size distribution is not modified inside the 
galactic wind and that the dust consists of silicate
grains. With these assumptions we 
calculated the dust injection rate in the same spherical model for
the cluster core as previously used (Table 1).
The results are given in Table 3.

The corresponding spectrum and brightness profile at 100\,${\mu}$m are given in
Figs. 2 and 3 with dashed lines. The amount of infrared emission produced 
by dust ejected by all early type galaxies in the inner core of Virgo 
cluster is negligible even in the upper limit calculation, being only at best
comparable to the emission from stellar dust.

\subsection{Dust originating in late-type galaxies}

Late-type galaxies are known to contain more dust than the elliptical galaxies,
and they can eject interstellar gas through
interaction with the cluster environment.  
Haynes \& Giovanelli (1986) have 
shown that Virgo spirals within 5 degrees of M87 are HI deficient compared
with their field counterparts. The HI deficiency becomes more marked for the
central regions, with
three-quarters of Virgo spirals within 2.5 degrees of M87
deficient by more than a factor 3 (${\rm DEF}> +0.47$). The summation of
HI deficiency by gas mass for spirals seen in projection
within 5 degrees of M87 (as tabulated by Giovanelli \& Haynes 1983 and
Haynes \& Giovanelli 1986) is 5.7$\times$10$^{10}\,$M$_{\odot}$, with the
bulk of the summed deficiency being due to the giant spirals.
Taking the dust (and gas) replenishment timescale in galactic disks
to be comparable to that estimated for the Milky Way of 
$\sim\,$3$\times$10$^{9}$yr
(e.g. Jones et al. 1997) one can estimate a current gas injection rate of 
$\sim\,$19$\,$M$_{\odot}$yr$^{-1}$ into the Virgo ICM 
from spiral galaxies. The dust content of the 
diffuse HI prior to ejection is difficult to 
estimate. On the one hand gas is preferentially lost from the outer disk,
where the metallicity and dust abundance is typically lower than that for the
galaxy as a whole.
On the other hand, it is known that spirals in the Virgo core  
have an enhanced metallicity by a
factor of approximately 2 compared with
field counterparts of similar lateness (Skillman et al. 1996). Assuming
that the ejected gas originally contained dust with an abundance of
0.0075 by mass (i.e. that of the Milky Way), 
we obtain a crude upper limit for the dust injection rate into the Virgo ICM 
of 0.14$\,$M$_{\odot}$yr$^{-1}$. This upper limit corresponds
to the case that grain sputtering 
in the injection process from the interstellar to the intracluster medium
can be ignored. 
It is somewhat lower that the 
estimated injection rate from the intracluster star population.

Whether the actual dust injection from spirals into the IC medium 
approaches this upper limit is likely to
depend on the mechanism for ejection. The most favoured model for
gas removal from the HI deficient galaxies, as originally proposed by 
Gunn \& Gott (1972), is ram-pressure stripping, whereby gas can be removed 
if the ram pressure P$_{\rm ram}\,\sim\,\rho_{\rm ICM}\,{\rm v}_{\rm gal}^
{2}$ exceeds the gravitational restoring force per unit area on the diffuse 
HI gas, P$_{\rm grav}=2\pi\,{\rm G}\,\sigma_{\rm grav}\sigma_{\rm HI}$. Here 
v$_{\rm gal}$ is the relative velocity of the galaxy through the IC medium of 
density $\rho_{\rm ICM}$, and $\sigma_{\rm grav}$, $\sigma_{\rm HI}$ are the 
gravitational and HI gas surface
densities in the galactic disk, respectively. 
Typically, giant spirals have 
$\sigma_{\rm HI}\,\sim\,10\,{\rm M}_{\odot}{\rm pc}^{-2}$ averaged over the
disk (e.g. Roberts \& Haynes 1994; Young \& Scoville 1991); the Milky Way
has $\sigma_{\rm HI}\,\sim\,5\,{\rm M}_{\odot}{\rm pc}^{-2}$ at the solar 
circle (Dickey \& Lockman 1990). Within 1$^{\circ}$ of M87,
$\rho_{\rm ICM}\,\geq\,\sim\,10^{-24}{\rm kg}\,{\rm m}^{-3}$, and typically
v$_{\rm gal}\,\sim\,1000\,{\rm km}\,{\rm s}^{-1}$. This is sufficient to strip
diffuse interstellar HI with 
$\sigma_{\rm HI}\,\leq\,10\,{\rm M}_{\odot}{\rm pc}^{-2}$ from within a 
giant spiral like the Milky Way, which has (Bahcall et al. 1992)
$\sigma_{\rm grav}\,\sim\,70\,{\rm M}_{\odot}{\rm pc}^{-2}$ at the solar 
circle.
To estimate whether grains
would be sputtered in this process 
we note that a sudden interaction of a galaxy with the IC medium would 
generally drive a shock wave into a cold
diffuse HI disk of density $\rho_{\rm HI}$ with speed
v$_{\rm s}\,\sim\,(\rho_{\rm ICM}/\rho_{\rm HI})^{0.5}\,{\rm v}_{\rm gal}$. 
For mid-plane number densities of $\sim\,0.5\,{\rm cm}^{-3}$, as in the 
diffuse HI in the
plane of the Milky Way at the solar circle (Dickey \& Lockman 1990),
v$_{\rm s}$ is 37$\,{\rm km}\,{\rm s}^{-1}$, for 
v$_{\rm gal}\,\sim\,1000\,{\rm km}\,{\rm s}^{-1}$ and 
$\rho_{\rm ICM}\,\sim\,10^{-24}{\rm kg}\,{\rm m}^{-3}$ (i.e. 1$^{\circ}$ from 
M87). This shock speed is well below the minimum value of 
$\sim\,100\,{\rm km}\,{\rm s}^{-1}$ needed 
for sputtering. Although sputtering should become effective in ram-pressure
stripping from the outer disk,
where HI number densities may be orders of magnitude lower, 
this simple consideration is consistent with the possibility that a 
substantial fraction of the
interstellar grains may survive the ram-pressure stripping process.
Indeed, some direct evidence for the presence of dust in 
interstellar material stripped from the Virgo elliptical galaxy M84 has been 
found by Rangarajan et al. (1995).

High resolution
hydrodynamical simulations of ram-pressure stripping of elliptical
galaxies (Balsara et al. 1994)
show that the gas removal involves a series of individual events
separated by $\tau_{\rm strip}$ of a few times 10$\,^{7}\,$yr, leading to
long tongues of stripped gas in the IC medium, a result which Balsara et al.
consider would also apply to gas stripping from spirals. Once injected into
the IC medium, the grains will be rapidly sputtered by the hot gas, so that
the infrared emission should still be morphologically
associated with the parent galaxy as a dust trail (Dwek et al. 1990),
rather than being smoothly distributed in the IC medium.
For the example
of an ambient number density of 0.0007$\,{\rm cm}^{-3}$ 
(corresponding to 1$^{\circ}$ from M87 - see Table 1), 
the sputtering time scale (Eq. (5)) is 
1.4$\times$10$^{8}\,$yr for grains of size 0.1${\mu}m$. 
The length of the infrared-emitting dust trail is then 
the distance traveled by the galaxy in this sputtering time, or 
146$\,$kpc for v$_{\rm gal}\,\sim\,1000\,$km$\,$s$^{-1}$.

We have estimated the infrared emission from a giant spiral with
a radius of 20\,kpc and an initial HI mass of 
about $6.6\times 10^9\,{\rm M}_{\odot}$. For a typical HI deficiency
(Haynes \& Giovanelli 1986)
of DEF=+0.47, the total HI mass loss is $4.4\times 10^9{\rm M}_{\odot}$.
Assuming most of the loss is sudden, on entering the cluster core region, 
and taking the
dust-to-gas ratio to be ${\rm Z}_{\rm d} = 0.0075$ (the Galactic value), 
then the mass of the dust released in the ICM medium by such a galaxy 
is ${\rm M}_d = 3.3\times 10^7\,{\rm M}_{\odot}$. 
The infrared spectrum of such a
source is given in Fig 4, for two cases; the first case (solid line) 
is when the stripping occurs at 10$^{\prime}$ from the cluster centre, 
(${\rm n}_{\rm e}=0.0039\,{\rm cm}^{-3}$, ${\rm T}_{\rm e} = 
2.36\times 10^7$\,K, see 
Table 1) and the second case (dashed-line) is when stripping occurs while the 
galaxy enters the dense core region of the cluster, at $1^{\circ}$ distance 
from the centre of the cluster (${\rm n}_{\rm e}=0.0007\,{\rm cm}^{-3}$, 
${\rm T}_{\rm e} = 3.37\times 10^7$\,K, see Table 1). It is interesting to note
that the IR flux density of the transient IR emission from the 
dust trail is predicted to rival that of the photon-heated dust in the 
galactic disk, and, despite the difference in heating mechanism,
have similar colours (with a spectral peak in the 100-200$\,{\mu}m$ range). 
This, combined with removal of dust from the disk, and hence reduced internal
extinction, will create a discrete system with brighter apparent blue 
magnitudes and a boosted spatially integrated IR flux density. If seen in a 
distant cluster, where the intracluster IR component could not be resolved from
the disk component, this could create the illusion of a galaxy with an 
enhanced star-formation activity, even though the star-formation in the
galaxy may actually be somewhat suppressed by the gas removal in reality.

\begin{figure}[htb]
\plotfiddle{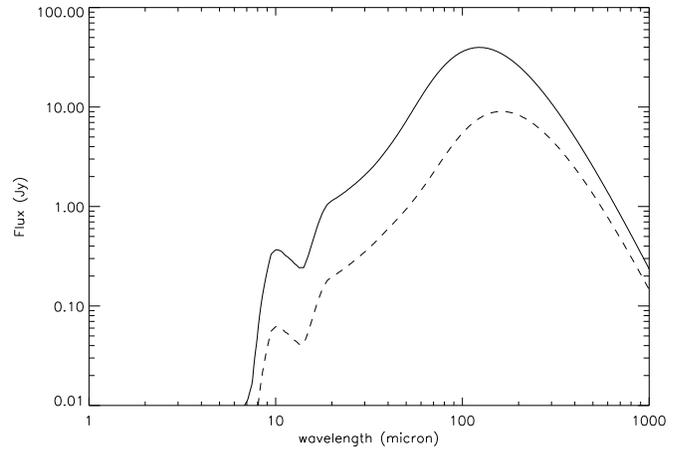}{3.0in}{0.}{50.}{50.}{-150}{-160}
\caption[]{The infrared spectrum of 
${\rm M}_{\rm d}=3.3\times 10^7\,{\rm M}_{\odot}$ of dust stripped from a 
spiral
galaxy while infalling into the dense core region of the cluster.
Solid line is when stripping occurs at  10$^{\prime}$ from the cluster centre, 
(${\rm n}_{\rm e}=0.0039\,{\rm cm}^{-3}$, ${\rm T}_{\rm e} = 
2.36\times 10^7$\,K, see 
Table 1) and dashed-line is when stripping occurs while the 
galaxy enters the dense core region of the cluster, at $1^{\circ}$ distance 
from the centre of the cluster (${\rm n}_{\rm e}=0.0007\,{\rm cm}^{-3}$, 
${\rm T}_{\rm e} = 3.37\times 10^7$\,K, see Table 1). 
A silicate dust composition was assumed. }
\end{figure} 

Since spirals are more spread over the whole cluster area, following the
density-morphological relation (Dressler 1980), they are by far less 
numerous in the centre of the Virgo cluster. From the number density profiles
of late-type galaxies in the Virgo Cluster and assuming a King profile with a 
core radius ${\rm r}_{\rm c}=3.2^{\circ}$ (Binggeli et al. 1987) we
estimate only a few spirals ($\sim 2$ galaxies) in the inner $1^{\circ}$ from 
the cluster centre. As ram-pressure stripping is likely to be a transient
phenomenon, occurring when the galaxies enter the dense core region of the 
IC medium, with sputtering timescales shorter than the time to 
traverse the angular extent of the core, we expect only of order 1 short-lived
intracluster IR source in the inner $1^{\circ}$ region of the cluster. 
To conclude, the IR emission coming from dust stripped from late-type 
galaxies is localised and connected to the 
parent galaxy, and does not account for a diffuse intracluster IR component.

\section{An infall model for the dust}

We have shown that the diffuse infrared emission from dust produced by discrete
sources inside the inner region of the Virgo cluster is below any detection
capabilities. This situation occurs from the fact that there are simply not
sufficient strong sources of dust to produce a dust injection rate high enough
to replenish the large amount of dust sputtered by the ambient hot plasma. 
In this section we propose a new mechanism for injecting dust into the ICM,
namely dust removed from spiral galaxies by galactic winds throughout
their life time, and brought into the cluster by the 
infalling spirals. This scenario is a corollary of the fact that the clusters 
are still in the process of forming, by accreting galaxies from the 
surrounding field environment. The Virgo Cluster is
a typical example (and best studied) for such a process. Tully \& Shaya (1984)
studied the phase space distribution of the spirals in and near the Virgo
cluster and developed a mass distribution model in which galaxies within about
8 Mpc of the Virgo cluster are now falling back towards the cluster. Within
their model most of the spirals entered the cluster in the last one-third to
one-half of the age of the universe. This implies that the Virgo spirals were
formed in lower density environments, more like field galaxies and have only
lately entered the high-density cluster environment. 

Following the idea of clusters accreting from the field, we propose the
following scenario (a schematic picture of this infall model is given in 
Fig. 5.):

1. Spiral galaxies are systematically infalling into the gravitational
potential well of the cluster. Following Tully \& Shaya (1984) we consider 
that all the spirals seen today in the Virgo Cluster have entered the cluster 
in the last 4-6 Gyr at a constant rate. Since the distribution of galaxies in 
the field is very clumpy, the infalling rate depends on the details of this 
distribution. But as we do not know the exact distribution of field 
galaxies over the last 4-6 Gyr, 
we can only assume a constant rate over this time. 
Furthermore, if spirals are destroyed or transformed into
 ellipticals when they 
approach the cluster core, then there may have been even more
spirals infalling into the cluster, than we see today. Thus our assumptions
give only a lower limit to the total number of infalling spirals. 

2. In addition to the infalling galaxies, there should,
presumably, be an accompanying primordial infalling diffuse intergalactic gas
component, from which the galaxies originally formed. Little is known about the
density and temperature of this intergalactic medium (IGM) in general, but 
it is believed to have temperatures in the range 10$^4-10^6$\,K. 
Immediately, prior to galaxy formation
one would expect the gas to co-move with the galaxies. At later stages of the
infall, gas-gas hydrodynamic interactions may however decouple the gas from the
galaxy inflow, and, provided the interactions are gentle enough to avoid
heating to virial temperature, would tend to increase the infall rate by
removing angular momentum. 

3. The diffuse low pressure ambient IGM is favourable to the formation of 
galactic winds in spirals \newline (Breitschwerdt et al. 1991). If we
assume that the galactic winds of spirals were able to inject dust into the IG
medium for their full lifetime of 13 Gyr, then large amount of dust is
also brought into the cluster. 

\begin{figure}[htb]
%\plotfiddle{unnamed.eps}{3.0in}{0.}{40.}{40.}{-120}{5}
\plotfiddle{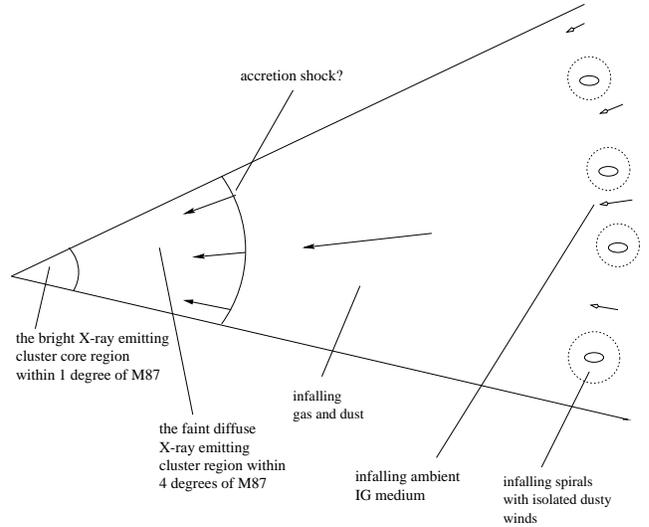}{3.0in}{0.}{40.}{40.}{-130}{5}
\caption[]{Schematic view of the infalling of gas and dust into the cluster
core }
\end{figure} 

A basic premise of our analysis is that
the spirals infalling into the cluster will be approximately comoving with 
the (primarily primordial) IG gas and the injected dust and gas
from the embedded
galaxies. This premise can only be indirectly tested as any infalling
IG medium can only be detected once it has been virialised into an 
X-ray emitting IC medium. 
Simple analytic models for the growth of clusters are given by
Gunn \& Gott (1972), in which the infalling baryonic material is
relatively cold, such that the hot X-ray emitting IC medium is separated from
the infalling gas by a strong accretion shock. 
In the present work, we tentatively identify the outer boundary of the
faint diffuse X-ray emission emission extending $4-5^{\circ}$ from M87 with
a macroscopic accretion shock.
The relatively
sharp radial cut off of the diffuse X-ray emission shown particularly by the
GINGA scan data (Takano et al. 1989) would seem to admit such an
interpretation. 
This allows us to estimate the current
upstream density of the infalling IG medium from the observed gas density
of $\sim\,4\times 10^{-5}\,{\rm cm}^{-3}$ (Schindler et al. 1999; see also 
Sect. 2) for the boundary of the diffuse X-ray emitting medium. Taking
the compression ratio of the accretion shock to be 4, this yields 
n$_{\rm IGM}\,\sim\,10^{-5}\,{\rm cm}^{-3}$ for the current number density of 
the infalling IG medium at a distance of 1.3$\,$Mpc from M87. 

We can now use this estimate for n$_{\rm IGM}$ 
in conjunction with the observed total
mass of baryonic matter emitting X-rays downstream of the accretion shock,
which should be made up of infallen matter and the initial perturbation, to
test the hypothesis that the infalling galaxies and IG gas are comoving.
We can estimate radial infall velocity of the IG gas at the radius of 
the presumed accretion shock of R$_{\rm shock}\sim1.27$\,Mpc (4$^{\circ}$) 
from the radial distribution of galaxian velocities derived by 
Tully \& Shaya (1984) in their simple radial infall model (see their Fig. 4). 
After taking into account the slight 
differences in the assumed distance to the cluster (16.8$\,$Mpc was derived 
by Tully \& Shaya) we predict the gas infall velocity at the accretion shock 
to be v$_{\rm infall}\sim\,800\,$km$\,$s$^{-1}$. From this, we can estimate the
current accretion rate of baryonic IG gas, assuming spherical symmetry, as
$4\pi\,$R$_{\rm shock}^{2}\,$m$_{\rm H}\,$v$_{\rm infall}\,$n$_{\rm IGM}\,
\sim$ \newline$3900\,$M$_{\odot}\,$yr$^{-1}$. Since both the IC medium and the infalling
galaxies show marked deviations from spherical symmetry, indicative of the
infall of an irregular distribution of clumps, this estimate should be treated
as a crude upper limit. Nevertheless, it is remarkable that the
3900\,M$_{\odot}\,$yr$^{-1}$ multiplied by a presumed age of 
$1.3\times10^{10}\,$yr, or $5.1\times10^{13}\,$M$_{\odot}$ is
close to the $4\times10^{13}\,$M$_{\odot}$ for the total baryonic
gas mass in the (dominant) IC medium component centered on M87 within
$1.3\,$Mpc, as plotted on Fig. 11a of Schindler et al. 
(1999). (Although
Schindler et al. use a hydrostatic model for the derivation of mass from
the X-ray emission, we would not expect very different results from a 
dynamic accretion flow model as the post shock inwards flow velocities 
are small). This result we use it here to justify our assumption of
a radial inflow of relatively cold IG gas, comoving
with the galaxies, which provides a medium to sweep
grains injected from the embedded spiral galaxies down into the cluster.
We also note that the incoming mass flux of the IG medium dominates that of the
incoming galaxies, so that it is a safe assumption that the
galactic winds themselves should have no significant heating
effect on the IG medium. 

Here we consider the fate of dust particles ejected in the winds of infalling
spiral galaxies. In the following we assume that grains always comove with the
ambient gas - i.e. that the grains are charged and the ambient gas is
magnetised. There are two main obstacles for such grains reaching the benign
environment of the IGM, which is too cool for significant sputtering. 

Firstly, the grains have to survive the passage through the hot galactic wind;
we will show in the next section that in most cases grains can survive this
passage. Secondly, the winds will create local \lq\lq bubbles\rq\rq\ in the
ambient IGM, containing wind shocks which provide a further
possibility for sputtering.  However, the densities are so small that this
effect is negligible. According to Castor et al. (1975), the density and
temperature in the hot shocked wind region is given by:

\begin{eqnarray}
{\rm n}_{\rm w} = 0.01\,{\rm n}_0^{19/35}({\rm \dot M}_6\,
{\rm v}_{2000}^2)^{6/35}\,{\rm t}_6^{-22/35}\,{\rm cm}^{-3}\\
{\rm T}_{\rm w} = 1.6\times 10^6\,{\rm n}_0^{2/35}({\rm \dot M}_6\,
v_{2000}^2)^{8/35}\,{\rm t}_6^{-6/35}\,{\rm K}
\end{eqnarray}
\\
where ${\rm \dot  M}_6 = {\rm \dot M}_{\rm w}/(10^{-6}\,{\rm M}_{\odot}\,
{\rm yr}^{-1})$, ${\rm v}_{2000} = {\rm v}_{\rm w}/(2000\,{\rm km}/{\rm s})$, 
${\rm t}_6 = {\rm t_{\rm gal}}/(10^6\,{\rm yr})$, M$_{\rm w}$ is the mass 
ejected through  the winds during the age of the galaxy t$_{\rm gal}$, at a 
constant rate ${\rm \dot M}_{\rm w}$, v$_{\rm w}$ is the terminal velocity of 
the wind, and n$_0$ is the number density of the
ambient IGM. For typical values of ${\rm \dot M}_{\rm w}=1\,{\rm M}_{\odot}\,
{\rm yr}^{-1}$, ${\rm v}_{\rm w} = 500$\,km/s 
(Breitschwerdt et al. 1991) and ${\rm n}_0 = 10^{-5}\,{\rm cm}^{-3}$ we obtain
${\rm n}_{\rm w}= 4 \times 10^{-7}\,{\rm cm}^{-3}$ and 
${\rm T}_{\rm w} = 2.1\times 10^6\,$K. 
For such a low density the sputtering time of a grain of size 0.1\,${\mu}$m is
$2.5 \times 10^{11}$\,yr and therefore this effect can be neglected.

A more fundamental effect is that the bubbles are buoyant, and will tend to 
rise relative to the cold ambient medium. This effect can potentially impede
the inflow of the grains. However, even if the bubbles could grow for a 
substantial fraction of the Hubble time, it seems very
unlikely that the velocity of a boyant bubble relative to the ambient medium
could become significant compared to the inflow velocities. A
fundamental upper limit for this relative velocity is the sound velocity
of the ambient medium, which for T$\sim10^5$\,K is $\sim 33\,{\rm km}\,
{\rm s}^{-1}$. This is much smaller than the expected escape velocity for a 
central mass of $2\times 10^{14}\,{\rm M}_{\odot}$ (Schindler et al. 1999), 
for the scales we consider here (out to 8$\,$Mpc). In particular we note 
that the Virgocentric infall of the Local Group is 220 km/s 
(Tammann \& Sandage 1985). 
In practice, as the galaxies approach the inner region of the cluster, they
will decouple from the velocity field of the IGM medium. Then it seems 
unlikely that the bubbles will remain intact, due to the expected relative
motions of the galaxies with respect to the primordial IGM. Analogous to stars
moving through the ISM of the Milky Way, the bubbles would be expected to 
develop axi-symmetric \lq\lq
cometary\rq\rq\ like structures, in which the hot material can escape into the
ambient medium along the trailing axis. We can use the estimate of the
density of the baryonic IG medium to check our assumption that sputtering 
can be neglected for grains released directly into the infalling IGM. 
Very conservatively, assuming that density varied with redshift as $(1+$z$)^3$
(i.e. ignoring the comoving density increase due to the infall into the 
cluster), n$_{\rm IGM}\sim10^{-5}\,{\rm cm}^{-3}$ corresponds to
$2.7\times10^{-4}$ at z$=2$, for which (using Eq. (19) and taking the
gas temperature to be constant in time at $10^{5}$\,K), the sputtering
timescale would be less than a Hubble time only for very small grains
(of size $\leq\,0.002{\mu}m$). Thus it seems reasonable to suppose that
grains can accumulate in the inflowing IG medium since rather early epochs. 
%This concludes our consideration
%of properties of the infalling IG medium having implications for the dust
%infall model.
%We make no further quantitative use of the derived value for 
%n$_{IGM}$ other than the use in the calculations of physical conditions in 
%expanding galactic wind bubbles embedded in the inflow in Sect. 4.

In summary, we would expect grains ejected in galactic winds to fall into the
cluster, co-moving with the ambient intergalactic gas. In the cold ambient 
gas outside the 
X-ray emitting region of the cluster the grains will survive, since there is 
no mechanism for destruction. But once the gas and grains reach the X-ray 
region, 
the grains will be sputtered and collisionally heated, and thus 
produce infrared emission. 

In order to  calculate the infrared emission from the intracluster medium, we 
must first consider destruction processes of grains in the galactic winds, 
which in principal could  reduce the grain injection rate.
 
\section{Grain sputtering and size distribution in the galactic winds}

From the study of the parameters of different kind of winds
(Breitschwerdt et al. 1991, Breitschwerdt \& Schmutzler 1999) we know that 
the initial conditions in the
winds are rather unfavourable to grain survival, due to the high densities and
temperatures. Thus the grain survival depends on how rapidly these winds can 
cool
and expand on their way out of the disk. There are mainly two types of galactic
winds in the spiral galaxies: winds driven by cosmic rays and
thermally driven winds (star-burst winds, like in M82). Cosmic
ray winds can be classified (Breitschwerdt \& Schmutzler 1999) in global 
winds, 
that originate from the large-scale expansion of the hot galactic corona, and 
local winds, which come from individual superbubble regions in the disk. 
Star-burst winds are stronger, being thermally driven due to the high gas 
temperature in their hot interstellar medium. 

Here we will consider the parameters of these 
different kinds of winds, as presented
in Breitschwerdt \& Schmutzler (1999), who included in their calculations 
both adiabatic and
radiative cooling. Especially for global winds it was shown that the radiative
cooling dominates the gas cooling until the temperature drops to a few 
$10^4$\,K,
when adiabatic cooling becomes important. Again, considering global winds 
Zirakashvili et al. (1996) have
shown that unsaturated non-linear Landau damping may dominate the advection of
waves in the plasma and hence lead to local dissipative heating. Up to now
there is no self consistent model for galactic winds that includes both the
effects of line cooling and heating due to wave damping. Below we argue that 
the contribution of wave heating is much smaller than that of cooling, and
can be neglected. Thus, we have calculated the cooling rate and the
heating rate using the solutions for global winds from 
Breitschwerdt \& Schmutzler (1999). The cooling rate C is:

\begin{eqnarray}
{\rm C} = {\rm n}^2\,\Lambda({\rm T})
\end{eqnarray}
\\
where n is the gas density in cm$^{-3}$ and the cooling function 
$\Lambda({\rm T})$ was taken from Kahn (1976), 

\begin{eqnarray}
\Lambda = 1.33\times 10^{-19}\,{\rm T}^{1/2} 
\,\,({\rm erg}\,{\rm cm}^3\,{\rm s}^{-1})
\end{eqnarray}
\\
The upper limit to the heating rate H is:

\begin{eqnarray}
{\rm H} = - {\rm v}_{\rm a} \nabla{\rm P}_{\rm c}
\end{eqnarray}
\\
with v$_{\rm a}$ the Alfv\'en velocity and P$_{\rm c}$ the cosmic ray pressure.
The effects of heating and cooling can be compared in Fig. 6. Above 1\,kpc 
from the disk, the cooling
rate is 3 orders of magnitudes higher, and it is still dominant at 10\,kpc. At
even higher distances the contribution of the two rates becomes equal, but the
gas temperature and density is already below the level where it could
produce any efficient dust grain destruction. Therefore we will consider in our
calculation the wind solution from Breitschwerdt \& Schmutzler (1999).

\begin{figure}[hbt]
\plotfiddle{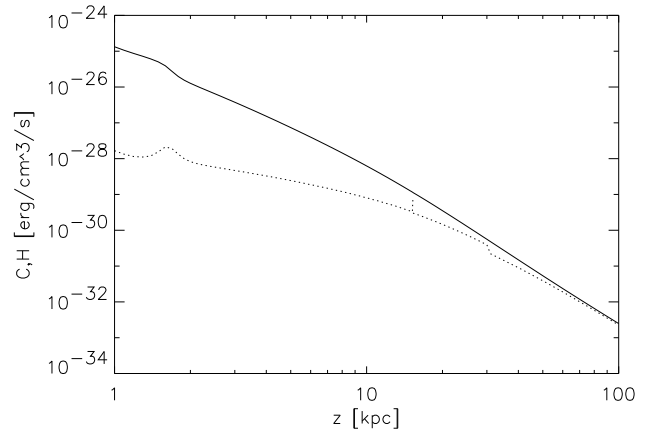}{3.0in}{0.}{45.}{45.}{-130}{-140}
\caption[]{The cooling rate C (solid line) and the wave damping
heating rate H as a function of the distance above the galactic disk d.}
\end{figure}

\begin{figure*}[hbt]
\plotfiddle{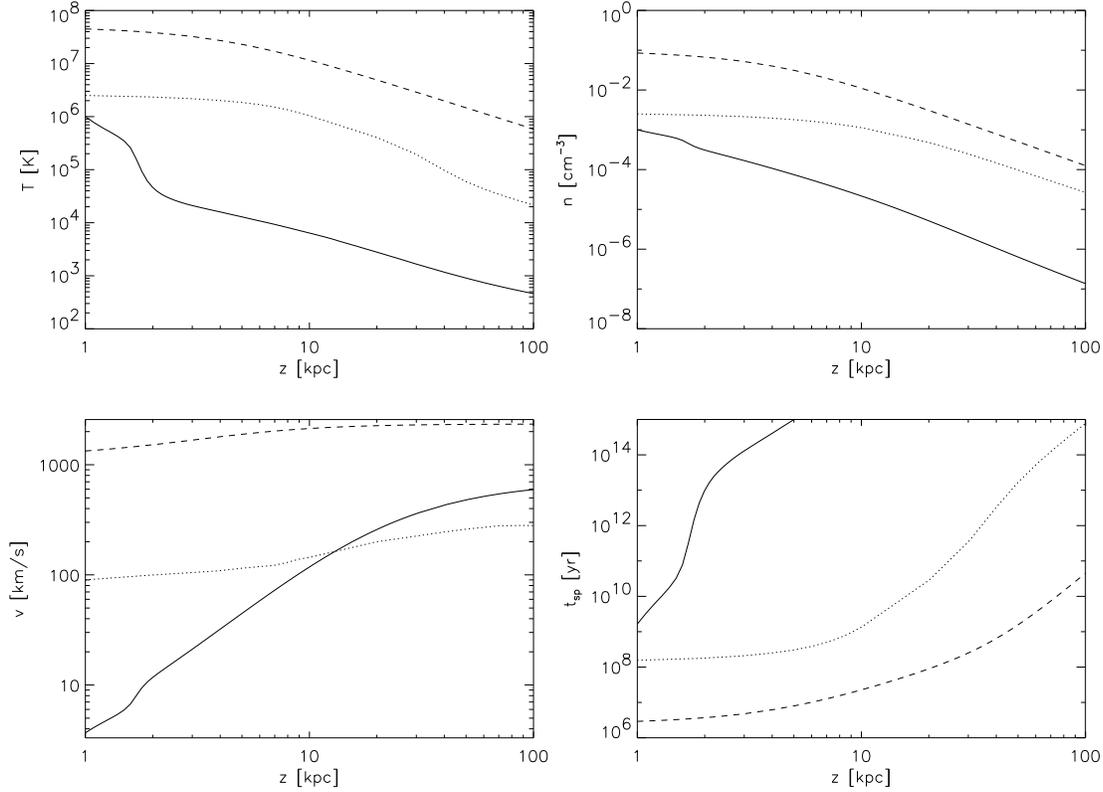}{4.5in}{0.}{80.}{80.}{-250}{-290}
\caption[]{The parameters of galactic winds: temperature T, density n, and
velocity v as a function of the distance above the galactic plane, z. The
sputtering time t$_{\rm sput}$ for a grain size of radius 
${\rm a}=0.25\,{\mu}$m is
also plotted in the lower-right panel. The global winds
are plotted with solid line, the local winds with short-dashed line, and the 
star-burst winds with long-dashed line.}
\end{figure*}

\subsection{Global winds}

In the slow global winds that are continuously emitted by the quiescent spiral
galaxies, the temperature drops very quickly with the expansion of the wind,
and this reduces substantially the grain sputtering.
The temperature (T\,[K]), density (n\,[cm$^{-3}]$), and velocity (v\,[km/s]) 
profiles above the disk (z[kpc]) are
plotted with solid line in Fig. 7, using the data from Breitschwerdt \& 
Schmutzler (1999) (also from D. Breitschwerdt, private communication). 
The global 
wind solutions correspond to a flux tube located at a galactocentric 
distance of 10 kpc, for an adiabatic model whereas an isochoric cooling 
function was included. These profiles were used for calculating the 
sputtering time profile, considering the analytical formula from Tsai \&
Mathews (1995). Thus, the rate at which the 
radius of the
dust grain decreases with time in a hot plasma of temperature T and density
n$_{\rm H}$ is:

\begin{eqnarray}
\frac{\rm da}{\rm dt} = -f\,{\rm n}_{\rm H}\,\left[\left(\frac{{\rm T}_{\rm d}}
{{\rm T}}\right)^{2.5}+ 1\right]^{-1}
\end{eqnarray}
\\
Tsai \& Mathews showed that this relation is a good approximation to the
detailed calculations of Draine \& Salpeter (1979) and Tielens et al. (1994),
for both graphite and silicate when 
$f=3.2\times10^{-18}{\rm cm}^4\,{\rm s}^{-1}$, and 
T$_{\rm d}=2\times 10^6$\,K. The
sputtering time can be then computed as:

\begin{eqnarray}
{\rm t}_{\rm sput} = {\rm a}\left |\frac{{\rm da}}{{\rm dt}}\right|^{-1}
\end{eqnarray}
\\
which for gas temperatures $10^6 < {\rm T} < 10^9$ K reduces to the formula 
(5). In Fig. 7 we show the sputtering time for a big grain of radius 
${\rm a}=0.25\,{\mu}$m. t$_{\rm sput}$ increases rapidly to more than 
$10^{10}$\,yr, 
but smaller grains have shorter survival times. The minimum grain 
size that can survive in the wind is derived from the cumulative size loss 
$\Delta$a, which is obtained by
integrating the sputtering rate over the time it takes the wind to go out the
galaxy: 

\begin{eqnarray}
\Delta {\rm a} = \int \frac{{\rm da}}{{\rm dt}} {\rm dt}
\end{eqnarray}
\\
The cumulative size loss $\Delta$a is plotted in Fig. 8 (solid line), and
from the saturation value we derive a minimum grain
size that survives sputtering ${\rm a}_{\rm surv}=0.0071\,{\mu}$m.

We assume that the initial size distribution in the wind is given again by a
power law of index k=3.5. The sputtering modifies the size distribution, in the
sense that grains smaller than a$_{\rm surv}$ will be completely destroyed while
bigger grains will end with a final size ${\rm a-a_{surv}}$. The final size
distribution will be given by:

\begin{eqnarray}
{\rm N}({\rm a}) \sim ({\rm a}+{\rm a}_{\rm surv})^{-k}
\end{eqnarray}
\\

In the case of global winds grains as small as $0.0072\,{\mu}$m can survive,
which means that the grain size distribution is only negligible changed, and
most of the dust will be injected in the IC medium. 

\subsection{Local winds}

Local winds have initially higher densities and temperatures, and these 
quantities remain almost constant till the wind reaches 10\,kpc above the 
galactic disk (Breitschwerdt \& Schmutzler 1999). The corresponding 
temperature, 
density and velocity profiles are plotted in Fig. 7 with short-dashed lines. We
expect that less grains will survive the wind. Following the same recipe like
in the case of global winds we obtain a minimum grain size that survives, ${\rm
a_{\rm surv}} = 0.084\,{\mu}$m. This is higher than the minimum grain size 
that survives the global winds, but still a substantial amount of
grains will survive. The corresponding sputtering time and cumulative size
loss ${\Delta}$a are plotted with short-dashed lines in Fig. 7 and 8,
respectively. 

\begin{figure}[hbt]
\plotfiddle{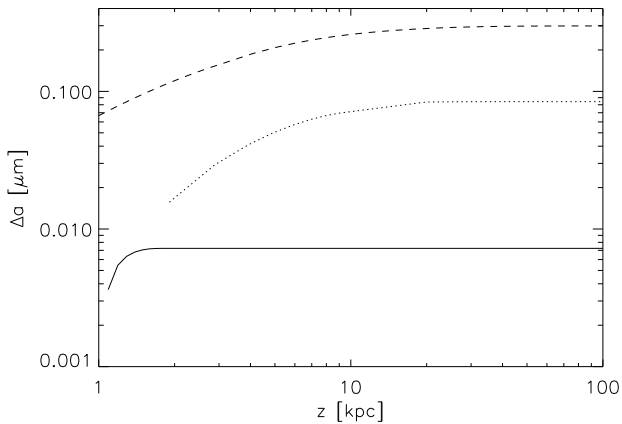}{3.0in}{0.}{45.}{45.}{-140}{-140}
\caption[]{The cumulative size loss ${\Delta}$a for different galactic winds,
as a function of the distance above the galactic plane, z. The global winds
are plotted with solid line, the local winds with short-dashed line, and the 
star-burst winds with long-dashed line. The saturation value for each 
distribution gives the minimum grain size that survives the wind.}
\end{figure}

\subsection{Star-burst winds}

The very energetic starburst winds can be traced already from 0.1\,kpc above
the disk, but in Fig. 7 and 8 we plot (long-dashed lines) only the range 
between 1 and 100\,kpc, in order to have the same range as for the other two 
type of winds. The temperature and density is more than one order of magnitude
higher than for global winds, and even at 10\,kpc these parameters
are still high enough to sputter quite efficiently the dust grains. 
Despite the fact that the wind velocity is also very high, the minimum grain 
size that
survives is ${\Delta}{\rm a}=0.3\,{\mu}$m. But such big grains
are rare in the diffuse interstellar medium. We conclude that star-burst winds
 are not able to inject dust into the IC 
medium. 

\section{The emission spectrum produced by dust falling into
the cluster core}

%\begin{figure}[htb]
%\plotfiddle{brightn_virgo_infall.ps}{3.0in}{0.}{50.}{50.}{-150}{-160}
%\caption[]{The infrared brightness profile at 175\,${\mu}$m produced by dust 
%infalling into the cluster core, in the ideal case that all the grains are
%sputtered at the outer boundary of the core. Only silicate grains were
%considered.}
%\end{figure} 

We have seen that dust grains survive mainly in global winds, and that their 
size distribution is not significantly changed by sputtering in the 
injection process. If the initial 
dust-to-gas ratio in the wind is Z$_{\rm d}=0.0075$ (the Galactic value), the 
grains will be injected into
the IG medium with Z$_{\rm d}=0.0071$ after sputtering, which means that the 
dust-to-gas
ratio is practically unchanged. The mass loss rate through global galactic
winds is 
of the order of $1\,{\rm M}_{\odot}/{\rm yr}$ for a galaxy like the Milky Way
(Breitschwerdt et al. 1991, Zirakashvili et al. 1996). The 
local winds can also inject 
grains into the IC medium, but with a smaller dust-to gas ratio, and with a 
modified size distribution given by (22). Furthermore, the mass loss
rate from these winds is one order of magnitude lower than for the global 
winds (Breitschwerdt \& Schmutzler 1999). And local winds do not happen
continuously over the life time of the spirals, but rather in episodes that are
connected with the life time of their parent giant HII region superbubbles. 
Thus we neglect
their contribution. Star-burst winds can inject
large amounts of gas in the IC medium, even more than 10\,M$_{\odot}$/yr, but
the injected gas is depleted of grains.
To conclude, spiral galaxies release grains in the IC medium mainly through
global galactic winds.

\subsection{Injection rate of grains into the intracluster medium}

We estimate the total infalling 
rate of grains into the cluster by
scaling the mass loss rate in galactic winds
to the blue luminosity of infalling spirals.
Following Tully \& Shaya (1984) (see Sect. 4) let us suppose that all galaxies 
currently seen within 6$^{\circ}$ of M87 arrived
at a constant rate within the last 5\,Gyr.\protect\footnote
{We note that this may actually yield a lower rate than the 
actual rate if some fraction of the infalling spirals had been transformed 
into ellipticals in this time.} These observed spirals have a 
total blue luminosity of $7.15\times 10^{11}\,{\rm L}_{\odot}$ 
(Binggeli et al. 1987), so that we can express
the infalling rate of luminosity as 
1.43$\times 10^{11}\,{\rm L}_{\odot}\,{\rm Gyr}^{-1}$.
We further assume that the galactic winds were able to inject dust into
the IG medium for their full lifetime of 13\,Gyr.
Then, scaling to the wind injection rate of 1\,M$_{\odot}\,$yr$^{-1}$ for 
the Milky Way (L$_{\rm B}\sim\,1.6\times10^{10}\,{\rm L}_{\odot}$;
de Vaucouleurs \& Pence 1978) as estimated by Breitschwerdt et al. (1991), 
each 
1.6$\times10^{10}\,{\rm L}_{\odot}$ entering the cluster will bring with it
an accumulated wind-ejected gas mass of 13$\times10^{9}\,{\rm M}_{\odot}$.
These estimates then lead to a gas infall rate from galactic winds of 
116\,M$_{\odot}\,$yr$^{-1}$. To calculate the corresponding grain infall rate
we take the dust-to-gas mass ratio in the ejected winds, averaged over
the 13\,Gyr, to be the solar value of 0.0075 to yield a dust injection
rate of 0.87\,M$_{\odot}\,$yr$^{-1}$. This may be quite conservative,
bearing in mind that the abundances of the galactic winds should be 
strongly enhanced compared to the ambient interstellar medium (V\"olk 1991).

In addition to dust injected into the external IGM by galaxies infalling
for the first time, there may also be a secondary
contribution from spirals which entered the hot IC medium at previous epochs, 
but have orbits taking them back beyond the accretion shock, where they 
can again inject dust.\footnote{We assume winds are suppressed
while the galaxies are in the pressurised IC medium.} The maximum
time since the currently seen spirals entered the cluster ($\sim\,$5\,Gyr) is
not much longer than the $\sim\,$3.2\,Gyr needed to traverse the 
volume enclose by the accretion shock on a radial path. Thus, only some 
fraction - perhaps one third - of the spirals will have passed back through 
the accretion shock. Their residence times after re-entry depends on the actual 
orbits and are thus very uncertain, but may be of the 
order of 2\,Gyr, in any case much less than the $\sim\,$13\,Gyr 
for the freshly infalling objects. If one third of the galaxies contribute
grains to the inflow for 2\,Gyr after reentry, the gas and dust injection
rates calculated above (for the freshly infalling systems) should be 
augmented by $\sim\,$5 percent to
122 and 0.91\,M$_{\odot}\,$yr$^{-1}$ respectively.

Finally, to calculate the dust injection rate into the
IC medium, we should take account of the fact that the accretion shock is
not stationary, but advancing outwards in the cluster reference frame
at some low speed v$_{\rm adv}$. Thus the true injection rate of grains into 
the hot IC medium will be somewhat greater than that given by the infall rate 
(as calculated above) by a factor 
$\zeta=$\,(v$_{\rm adv}+$v$_{\rm infall}$)/v$_{\rm infall}$. A rigorous
calculation of $\zeta$ is beyond the scope of this paper, but an
upper limit - for the case that the accretion shock had a compression ratio of
4 and the material downstream of the accretion shock was 
almost stationary in the cluster reference frame - would be 1.33.\footnote{In 
reality the downstream flow clearly has to have some inwards velocity to
maintain pressure equilibrium against gravity in the IC medium.} Another
estimate for $\zeta$ can be obtained from the estimate of
v$_{\rm infall}\sim800$\,km$\,$s$^{-1}$ from the infall of the galaxies
(following Tully \& Shaya (1984) - see Sect. 4) and a crude value for 
v$_{\rm adv}$ of $\sim800$\,km$\,$s$^{-1}$ which is the accretion shock 
radius (1.27\,Mpc) divided by the age of the cluster (13\,Gyr). 
This yields $\zeta\leq$1.125. We will assume $\zeta=$1.1, from which
we obtain final estimates of 134 and 1.0\,M$_{\odot}\,$yr$^{-1}$ for the
gas and dust injection rates into the IC medium.

\subsection{Heating and sputtering of injected grains}

In order to calculate the heating of grains injected from the
inflowing IG medium we recall consider the X-ray
morphology of the Virgo Cluster, as summarised in Sect. 2.
We only consider the dominant M87 subcluster, which accounts for
71$\%$ of the total emission out to 4-5$^{\circ}$.
This consists of the dense hot X-ray core region 
extending 1$^{\circ}$ around M87, and 
the faint diffuse emission extending 4-5$^{\circ}$ from M87, which
produces 15$\%$ of the total X-ray flux (B\"ohringer et al. 1994). We have
identified the boundary of the faint diffuse emission as an accretion shock 
(Sect. 4).

Grains injected at the supposed accretion shock should be heated by a plasma of
density $\sim\,4\times 10^{-5}\,{\rm cm}^{-3}$ and temperature 
$\sim 3\times10^{7}$K. The sputtering timescales in
this most tenuous region of the IC medium are 2.5$\times10^{9}$\,yr of grains
of size 0.1\,${\mu}m$.
Although comparable to the infall time for galaxies from the 
accretion shock radius to the centre, this sputtering timescale does not 
mean that the IR emission will be distributed over a broad range of radii.
The grains will follow the
inflow of the gas downstream of the accretion shock, which should be actually
quasi hydrostatic. Thus, in this simple homogeneous picture for the cluster 
the grains would be expected to be heated and sputtered in the 
vicinity of the accretion shock 4-5 degrees from the cluster core, and the 
infrared emission would trace the morphology of the accretion shock surface.
In this scenario, the grain heating will be relatively low as the grains
never reach the denser core regions of the IC medium, and the emission will
predominantly arise in the submillimetre range. We refer to this as case A.
 
In reality the structure of the M87 subcluster is much more 
irregular than this simple representation. B\"ohringer et al. (1994) 
have shown that in the western part 
of the cluster the X-ray emission falls off more steeply, while the northern 
edge of the cluster is less well defined, with the cluster boundary 
dissolving into several individual subclumps. 
Because of the uncertainties in the 3D structure of the cluster, it is very
difficult to predict the actual fate of infalling grains reaching the 
accretion shock radius. In particular, the accretion may be fundamentally 
clumpy in nature, with infalling clumps of dust-bearing gas reaching 
different depths before interacting and merging with an 
irregularly shaped intracluster medium. In this scenario it
may be possible for clumps infalling through certain position angles to 
directly interact with the dense X-ray core region extended 1$^{\circ}$
around M87. Then the grain heating will be stronger, with most radiation 
being produced in the FIR. We refer to this as case B.

\begin{figure}[htb]
\plotfiddle{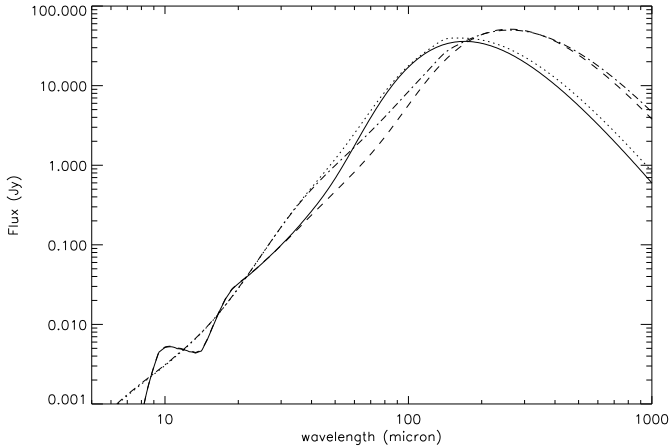}{3.0in}{0.}{50.}{50.}{-150}{-160}
\caption[]{The infrared spectrum produced by dust infalling into the cluster 
core, in two extreme cases: the grains are mainly heated by the diffuse X-ray
plasma inside the 4$^{\circ}$ radius region (dashed-line for silicate
grains and dashed-dotted lines for graphites); the grains are heated when 
they enter the dense 1$^{\circ}$ core region of the Virgo cluster 
(solid line for silicate grains and dotted lines for graphite grains).
 }
\end{figure} 

We note that the instantaneous luminosity will be very similar which ever of
case A and B is nearer the truth, provided the grain injection rate is 
constant over the sputtering timescale. For a steady state between
injection and destruction, the effect of the increased heating expected
for grains reaching the central dense regions of the IC medium (Case B) will
almost exactly be balanced by the shorter grain survival times, since 
sputtering timescales are only weakly dependent on temperature for plasma
hotter than $\sim\,$10$^{6}\,$K.

The spectrum of the submillimeter emission 
in case A is given in Fig. 9,
both for silicate grains (dashed line) and for graphite grains 
(dashed-dotted line). In the calculation we assume a steady-state solution for
the balance between dust destruction and injection just downstream of the
accretion shock, following the procedure described in 
Sect. 2. The peak of the emission, the colour 
of the spectrum and the fluxes are given in Table 4. 
Fig. 9 also shows the predicted emission from case B (solid line for 
silicates and dotted line for graphites). Here the spectrum was calculated 
under the assumption that the grains will
survive and emit only in the outer spherical shell of the central
core of cluster at radii of $\sim\,$1$^{\circ}$ 
(n$_{\rm e}=0.0007\,{\rm cm}^{-3}$, T$_{\rm e}=3.37\times 10^7$, see Table
1). The parameters of this emission are also given in Table 4.

\begin{table}[htb]
\caption{The characteristics of the infrared spectrum in two extreme cases (A
and B), both for silicate and graphite grains. The table lists the wavelength
corresponding to the maximum emission, the colour of the spectrum and fluxes.}
\begin{tabular}{|l|l|l|l|l|l|l|}
\hline\hline
 & \multicolumn{3}{c|} {Case A (${\rm r}=4^{\circ}$)} & \multicolumn{3}{c|} 
{Case B (${\rm r}=1^{\circ}$)}\\
\hline
& & & & & &\\
& ${\lambda}_{max}$ & $\frac{{\rm F}{259}}{{\rm F}{175}}$ & F$_{259}$ & 
${\lambda}_{max}$ & $\frac{{\rm F}{175}}{{\rm F}{100}}$ & F$_{175}$\\
& [${\mu}$m] & & [Jy] & [${\mu}$m] & & [Jy] \\
& & & & & &\\
\hline
   & & & & & &\\
Si & 259 & 1.4 & 51.3 & 172 & 2.1 & 35.9 \\
Gra & 259 & 1.3 & 50.0 & 163 & 2.2 & 39.4 \\
   & & & & & &\\
\hline\hline
\end{tabular}
\end{table}

\section{Discussion}

We have calculated the infrared emission for the Virgo 
cluster taking into account all 
possible sources of dust inside the cluster core. We have shown that
there are not enough discrete sources of dust in the Virgo cluster to produce 
detectable diffuse emission. This is compounded by the fact that galaxian
sources of dust embedded in the IC medium will certainly not be able to
provide a smooth distribution of grains within the volume of the IC medium as
sputtering timescales are invariably shorter than transport timescales
over a typical separation between galaxies. Of all potential discrete
sources only the stars could be thought of as giving rise to a smooth
emission component, though, for the Virgo cluster, this will only amount
to 20$\,$percent of the emission from the infalling intergalactic grains.
Thus, any detection of truly diffuse FIR emission is likely not to trace 
dust injected by galaxies inside the cluster core, but rather the inflow of 
grains to the
cluster from the external intergalactic medium.
Since younger clusters, like
the Virgo cluster still have spiral galaxies infalling into the cluster, we
might expect them to have a larger amount of infalling dust, and thus a higher 
FIR emission. Thus, in general, the phenomenon of diffuse intracluster FIR 
emission may give information on the current dynamic age of the cluster.
In this sense it is
complementary to the measurements of X-ray emission which broadly relate to
all the baryonic matter accumulated over the lifetime of the cluster.

\subsection{Detectability and recognition of IC dust emission}

Our estimates for the IR emission rely on a chain of quite poorly
determined quantities, such as the dependence of mass ejection in galactic 
winds on blue luminosity, uncertainties in the dynamical properties of the 
inflow, and the metallicity of the wind ejected material. Detection of diffuse
IR emission may be useful to constrain some of these factors. On the basis of
the simple estimates given in Table 4, however, it will be fundamentally
difficult to detect such emission from nearby clusters. The high angular sizes
will lead to severe confusion of the IC IR emission with foreground emission, 
which will be difficult to alleviate with multiwavelength studies given that 
the predicted emission has quite similar colours to the galactic cirrus. 
The best possibility for detection is for our case B scenario in which the 
cluster
is fundamentally clumpy, as then the emission may be concentrated over 
smaller solid angles (scales of 1$^{\circ}$ as opposed to 4$^{\circ}$ 
for the case A scenario). In general, because the grain sputtering timescales 
are shorter than the grain transport timescales, once the grains have 
reached their sputtering/injection sites, they will in general trace 
surfaces of injection rather then be distributed over a depth of volume.
This may lead to some limb-brightening effects which could aid detection,
though care would have to be taken to distinguish these structures from the
trails predicted from spiral galaxies undergoing ram-pressure stripping
discussed in Sect. 3.3.

It is more likely that diffuse intracluster emission could be
detected towards distant clusters of lower angular size. Referring again
to our illustrative scenarios of grains being sputtered in the 
diffuse X-ray emitting plasma (Case A) or when they first enter the 1 degree 
core region of the cluster (Case B), we address the question of how  
the Virgo cluster would appear if situated at a cosmological distance of 
z=0.5. The 4 degree region 
of the cluster would be seen as a source of 3$^{\prime}$. So the cluster will 
look like a compact source, and to detect this source we need to have only 
enough total flux. If we consider a Euclidian flat geometry with q$_0$=1/2 
we obtain a flux of 1.5 mJy (redshifted at ${\lambda}=389\,{\mu}$m) for Case A
or a flux of 1 mJy (redshifted at ${\lambda}=263\,{\mu}$m) for Case B. There
will be of course the emission from dust within the cluster galaxies. Here,
the main confusion problem would be with the FIR emission from constituent
spiral galaxies. To 
estimate the galaxy contribution to the total FIR emission we used the 
correlation of the B magnitude of 105 disk galaxies selected from the Virgo
Cluster Catalogue with 
their IRAS 100 micron emission. We found S$_{\nu}(100{\mu}m) 
\sim {\rm S}_{\nu}({\rm B})^{1.25}$ in mJy. Using again a King profile for 
the Virgo Cluster 
spiral galaxies we derived a total blue luminosity within the 4 degrees 
region from the cluster centre of 
L$_{\rm B}= 3.7\times 10^{11}\,{\rm L}_{\odot}$. 
This translates into a flux of 295\,Jy at 100 micron. For a colour 
F$_{175}/{\rm F}_{100}=1.5$ we estimate that the cluster galaxies should 
contribute with 442\,Jy to the total flux at 160\,${\mu}$m. This translates 
into a flux of 13\,mJy for a redshifted cluster, which is
still one order of magnitude higher than the intracluster emission. Thus,
a combination of good surface brightness performance and resolution will be
required for a unambiguous detection of IC dust emission. Especially
as the emission peaks in the submillimeter, this type
of observation will be particularly well suited to the new generation of
submillimeter interferometers.

Finally, we remark that galaxies may not be the only sources of IG grains,
especially in the early universe. FIR/ sub-mm observations of clusters may in
general provide new information of the abundance of the IG grains, which is
currently only poorly constrained.

\subsection{The optical extinction in clusters}

From our infalling model it is obvious that the extinction in clusters will be
dominated by dust upstream the accretion shock. If we suppose that grains are
uniformly distributed within the volume of the cluster, we obtain an 
A$_{\rm B}= 0.005$ magnitudes at a distance 4 degrees from the cluster center.
 However, this extinction could increase when some limb brightness effects 
are taken into account. Furthermore, extinction may be higher on some optical 
paths if we allow for clumpy accretion of infallen material. In passing we
mention that these calculations are obtained under our very conservative assumption
that galactic winds have the Galactic gas-to-dust ratio. As discussed in
Sect. 6.1, the abundance of galactic winds may be strongly enhanced as compared
to the ambient interstellar medium (V\"olk 1991) and thus much more dust can
then be injected in the IG medium. Thus the amount of extinction may be
sensitive to the assumed accretion rate of dust and to the geometrical 
effects in
the cluster. Our predicted extinction should be compared to the few tenths of 
optical depth predicted by some optical studies which found a deficit of 
distant galaxies (Zwicky 1962; Karachentsev \& Lipovetskii 1969; 
Bogart \& Wagoner 1973; Szalay et al. 1989) or quasars (Boyle et al. 1988; 
Romani \& Maoz 1992). However there are also 
some optical studies that do not support the presence of extinction in 
clusters. Maoz (1995) found out that radio-selected
quasars behind Abell clusters are not redder than quasars that do not have a
cluster in foreground. He also suggested that previous results that
claimed the avoidance of foreground Abell clusters by optically selected
quasars can be explained as being due to selection effects. Ferguson (1993)
studied the colour versus Mg$_2$ relation for elliptical galaxies in cluster,
group and field samples. He also found no evidence for excess reddening in 
clusters or groups. 

\subsection{Relation to other observables}

Our predictions for diffuse FIR/sub-mm intracluster emission rely on 
the ubiquitous existence of ``global'' winds from quiescent spiral
galaxies, enriching the infalling intergalactic medium with metals,
partly as dust grains.
As noted by V\"olk (1991), it should be expected that the metallicity
of the ejected wind material should be substantially higher than for
the interstellar medium of the disks, which has consequences for the
chemical evolution of spirals. An interesting corollary, therefore, of
any detection of diffuse IR emission in the periphery of the cluster,
is the implication for the environmental effects on the chemical 
evolution of spiral galaxies. 
There is some evidence (Skillman et al. 1996) that spiral galaxies in the
core region of the Virgo cluster have enhanced metallicities compared to
counterparts in the cluster periphery and field environment. 
Skillman et al. (1996) account for this abundance contrast 
in terms of scenarios invoking the accretion of
metal-poor cold intergalactic material to dilute the abundances of field
galaxies, by supposing that the accretion is suppressed for galaxies in the 
cluster core due to the hot environment.
We propose the opposite scenario to explain the abundance contrast between
field and cluster, namely the ejection of overabundant interstellar
material in quiescent winds of field galaxies, as opposed to the
cluster core, where quiescent winds are predicted
to be suppressed by the high pressure environment.
We note that a wind scenario for chemical abundance regulation would
require a larger integrated nucleosynthetic production over the
lifetime of a galaxy as compared to a regulation by accretion, due to the 
loss of a substantial amount of metals to the IG medium.

The infall scenario may also alleviate the problem of explaining the 
absolute metal content of
the intracluster medium purely in terms of the nucleosynthetic 
history of galaxies, principally ellipticals (e.g. Okazaki et al. 1993) 
embedded in the IC medium, as some proportion of the metals
will then have been injected through the winds of spirals
embedded in the infalling IG medium. Recently, Wiebe et al. (1999) also 
considered the possibility that galactic winds from spiral 
galaxies contribute to the chemical enrichment of the ICM, though they did 
not take into account the fact that winds are suppressed in the dense ICM by 
the high pressure environment. In our infall model, one possible observational
manifestation would be the expected X-ray emission
in the periphery region of the cluster, just downstream of the accretion 
shock. Once sputtered, the grains will release refractory elements into the
gas phase, which will radiate line emission to supplement emission
from the directly injected non-refractory component.
The radial profiles of line emission may give some indication of the infall 
history of these metals. 

Our predictions for the origins of diffuse infrared emission in the IC medium
may also have a physical corollary to the observational situation regarding the
diffuse synchrotron emission. Grains are short-lived and cannot move far from
their sources. But by the same token, relativistic electron cannot move far 
from their sources due to inverse Compton and synchrotron losses. It is 
tempting to imagine a scenario in which both phenomena are localised around 
an accretion shock and would be interesting to compare the morphology of 
diffuse IR and radio emission from this viewpoint. The possibility of particle
acceleration at the location of extended accretion shocks around the clusters
has been recently discussed by En{\ss}lin et
al. (1998). They identified the so called cluster radio relics to be powered by
the accretion shock produced by the large-scale gas motion around the clusters.

Finally, we can examine whether submillimeter emission
from intracluster dust, as predicted here, might be detectable 
in experiments to detect Sunyaev-Zeldovich (SZ) excess towards
clusters. At increasing redshifts, the wavelength of peak emission
from the IC dust may move close to the SZ peak around $\lambda=0.8$\,mm.
The appearance of both the intracluster dust emission and the SZ
excess is predicted to be related to the morphology of the diffuse X-ray
Bremsstrahlung from the ICM. To get an estimate for the relative
emissions from the two phenomena we have estimated the (thermal)
SZ effect that we would observe for a cluster
like the Virgo Cluster, using the parameters for the 4 degree central 
region of Virgo, but with the cluster redshifted to z=0.5. The Comptonization 
parameter y (see Rephaeli 1995) was calculated by integrating the
gas density profile of the cluster (Schindler et al. 1999) for a constant 
temperature
${\rm T}=3.3\times 10^7$\,K. We obtained an averaged 
${\rm y}=4.3\times 10^{-6}$, which is consistent
with the upper limit determined from the COBE/FIRAS database, 
${\rm y}\leq 2.5\times 10^{-5}$ (Mather et al. 1994). The corresponding
temperature change due to the scattering was calculated in the non-relativistic
case (Rephaeli 1995):

\begin{eqnarray}
\Delta T_{\rm nr} = \left [ \frac{x\,(e^x+1)}{e^x-1}-4\right]\,{\rm T}_0\,{\rm y}
\end{eqnarray}
\\
where x=h${\nu}$/kT is the nondimensional frequency, T is the radiation
temperature and ${\rm T}_0=2.726$ (as determined by the COBE/FIRAS; Mather et
al. 1994). The corresponding surface brightness due to this change in
temperature was multiplied with a solid angle of $3^{\prime}$ radius (as
seen for the 4 degrees inner region of the Virgo Cluster at a 
redshift z=0.5) and the
integrated flux is shown in Fig. 10. The fluxes estimated for the diffuse dust
emission (in both Case A and Case B) for the Virgo Cluster (redshifted to
z=0.5) are also plotted for comparison in Fig. 10. It is obvious that for 
wavelengths less than 400 micron the diffuse emission is dominated by dust 
emission. At longer wavelengths the dust emission becomes less important, 
reaching about 5 percent of the S-Z emission at the S-Z peak.
However, these considerations relate to the integrated fluxes. 
If the diffuse dust emission could be spatially resolved, as indeed it must
be to distinguish it from emission from discrete sources in the cluster,
it may be possible to distinguish it from the SZ emission due to it's 
predicted limb brightening at the boundary of the accretion shock. Indeed,
in the outermost regions of the cluster, IC dust emission may
give a significant signal compared to the smoothly, centrally peaked SZ 
emission. 

\begin{figure}[htb]
\plotfiddle{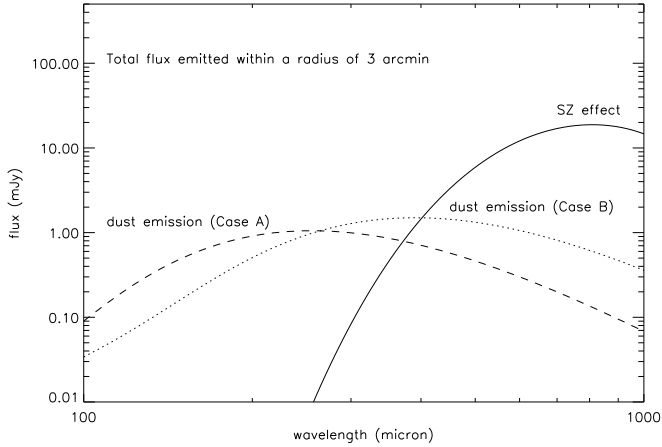}{3.0in}{}{50.}{50.}{-150}{-160}
\caption[]{Predicted flux emitted within a radius of 3 arcmin by a cluster like
Virgo Cluster, at a redshift z=0.5. With solid line we plot the 
thermal SZ effect, with
dashed line the dust emission for Case A and with dotted line the dust emission
as estimated for Case B.}
\end{figure} 

\section{Summary}

In this work we show that any detection of a diffuse FIR intracluster emission
is likely not to trace dust injected by galaxies inside the cluster
core, since there are not enough discrete sources of dust to produce
detectable emission. We propose that the diffuse intracluster FIR emission may
trace the current accretion rate of the cluster, which give information on the
current dynamic age of the cluster. Even in this case the estimated amount of 
dust in the cluster is lower than suggested in earlier work (Dwek et al. 1990). 
The main results of this paper are listed 
below:
\begin{itemize}
\item 
The infrared emission from dust injected by intergalactic stars inside the core
region of the Virgo cluster is a factor of $\sim 10$ below the detection limit
of currently available telescopes.
\item The amount of intracluster infrared emission produced by dust ejected by
early type galaxies in the inner core of Virgo cluster is negligible even in
the upper limit calculation.
\item The IR emission coming from dust stripped from late-type galaxies is
localised and connected to the parent galaxy, and does not account for a
diffuse intracluster IR component.
\item
For dynamically young clusters like the Virgo cluster we propose a new
mechanism for injecting dust in the ICM, namely dust removed from spiral
galaxies by galactic winds throughout their life time, and brought into the
cluster by the IGM. A basic premise of this scenario is that the
spirals infalling into the cluster comove with their ambient IG medium and the
injected dust and gas from the embedded spirals.
\item In this work we identify the outer boundary of the diffuse X-ray 
emission of the Virgo Cluster, extending $4-5^{\circ}$ from M87, with a
macroscopic accretion shock. In a simple homogeneous picture for the cluster
the infalling grains would trace the morphology of the accretion shock
surface. Since the density of the ambient plasma is low in this region, the
heating is relatively low and the infrared emission arise in the submillimetre
range. If the accretion is fundamentally clumpy in nature, it is possible for
clumps infalling through certain position angles to directly interact with the
dense X-ray core region of the cluster. Then the infrared emission is shifted
to shorter wavelengths.
\item
For nearby clusters it will be fundamentally difficult to detect infrared
emission from intracluster dust, due the large volumes considered and to the
relative low masses of dust. It is more likely that diffuse intracluster 
emission could be detected for distant clusters. Because the cluster emission 
is dominated by emission from cluster member galaxies, any detection of a 
diffuse component requires a combination of good surface brightness 
performance and resolution. This type of observation would be well suited to 
the new generation of submillimeter interferometers. 
\end{itemize}

\acknowledgements

We would like to thank Dr. M. Voit for the careful refereeing of the manuscript.
We gratefully acknowledge Dr. D. Breitschwerdt for providing us with his 
calculated data on galactic winds. We also kindly acknowledge Dr. U. Fritze-von
Alvensleben for providing us with the HRD population model needed in Sect.
3.1. We would like to thank Drs. A. Burkert and A. Blanchard for interesting 
discussions.

\end{document}